%% file: main.tex
\documentclass[11pt,a4paper]{article}
\baselineskip
\pdfoutput=1

\newcommand{\sech}{\mathrm{sech}}
\newcommand{\eq}{\begin{equation}}
\newcommand{\en}{\end{equation}}

\usepackage[utf8]{inputenc}
\usepackage[english]{babel}
\usepackage[a4paper, margin=3.0cm]{geometry}
\usepackage{amsmath}
\usepackage{amssymb}
\usepackage{pdfpages}
\usepackage{comment}

\usepackage{orcidlink}

\usepackage{titlesec}
\usepackage{multirow}
\usepackage{adjustbox}
\usepackage{subcaption}
\usepackage{graphicx}
\usepackage{array}
\usepackage{hyperref}
\usepackage{cite}
\usepackage{booktabs}
\usepackage{float}
\usepackage{adjustbox}
\usepackage{braket}
\usepackage{placeins}

\usepackage{tikz}
\usetikzlibrary{arrows.meta}
\usetikzlibrary{positioning}
\usetikzlibrary{decorations}
\usetikzlibrary{decorations.markings}

\usepackage{graphicx} 

\DeclareMathOperator{\Tr}{\mathrm{Tr}}
\DeclareMathOperator{\diff}{\mathrm{d} \!}
\DeclareMathOperator{\SU}{\mathrm{SU}}

\DeclareMathOperator{\erfc}{\mathrm{erfc}}
\newcommand{\real}{{\rm Re\,}}

\newcommand{\intr}{\mathrm{intr}}
\newcommand{\NambuG}{Nambu-Got\={o}\;}
\newcommand{\ndof}{n_{\mathrm{dof}}}

\begin{document}
\begin{titlepage}
\renewcommand\thefootnote{\mbox{$\fnsymbol{footnote}$}}
\begin{center}
\Large\bf Intrinsic Width of the Flux Tube as a tool to explore confining mechanisms in Lattice Gauge Theories

\end{center}
\vskip1.3cm
\centerline{
Michele~Caselle\orcidlink{0000-0003-3115-1170}$^{1}$\footnote{\href{mailto:michele.caselle@unito.it}{{\tt michele.caselle@unito.it}}},
Elia~Cellini\orcidlink{0000-0002-5664-9752}$^{1,2}$\footnote{\href{mailto:elia.cellini@unito.it}{{\tt elia.cellini@ed.ac.uk}}},
Alessandro~Nada\orcidlink{0000-0002-1766-5186}$^{1}$\footnote{\href{mailto:alessandro.nada@unito.it}{{\tt alessandro.nada@unito.it}}},
Dario~Panfalone\orcidlink{0009-0007-6651-7490}$^{1}$\footnote{\href{mailto:dario.panfalone@unito.it}{{\tt dario.panfalone@unito.it}}}
and Lorenzo~Verzichelli\orcidlink{0009-0008-0825-4845}$^{1}$\footnote{\href{mailto:lorenzo.verzichelli@unito.it}{{\tt lorenzo.verzichelli@unito.it}}}
}

\vskip1.2cm

\centerline{\sl $^{1}$Department of Physics, University of Turin and INFN, Turin}
\centerline{\sl Via Pietro Giuria 1, I-10125 Turin, Italy}

\centerline{\sl $^{2}$Higgs Centre for Theoretical Physics, School of Physics and Astronomy}
\centerline{\sl The University of Edinburgh, Edinburgh EH9 3FD, United Kingdom}

\vskip1.0cm

\setcounter{footnote}{0}
\renewcommand\thefootnote{\mbox{\arabic{footnote}}}
\begin{abstract}
\noindent
We study the profile of the flux tube in the $\mathrm{SU}(2)$ gauge model in $(2+1)$ dimensions, with a particular attention to the so called ``intrinsic width'' which drives the exponential decay of the flux density at large transverse distances. This quantity is directly related to the confining mechanism which generates the flux tube: to test the properties of the latter we study a wide range of different values of lattice spacing, temperature and flux tube lengths and show that our data are precise enough to distinguish between different confining models.  In particular we show that at high temperatures (just below the deconfinement transition) the data are perfectly described by an Ising-like effective model based on the Svetitsky--Yaffe mapping.  At lower temperatures this approximation does not hold anymore. In this regime (which is the most interesting one from a physical point of view) we test several alternative proposals and show that the dual superconductor model is the one which better fits the data. However, this proposal is not fully satisfactory, because the values of the Ginzburg--Landau parameter extracted from the fits increase with the length of the flux tube, which is not a feature predicted by the model. This suggests that a more sophisticated model is needed to explain confinement in non-abelian gauge theories and, at the same time, that our data on the intrinsic width may be a powerful tool to benchmark these candidates. 
\end{abstract}

\end{titlepage}

\tableofcontents

\section{Introduction}

A defining signature of confinement in non-abelian gauge theories is the formation of a flux tube connecting two static colour sources. Within the Effective String Theory (EST) framework, this flux tube is modelled as a fluctuating string joining the quark and the antiquark~\cite{Luscher:1980ac, Luscher:1980fr}. Although this string is assumed to be infinitely thin, quantum fluctuations generate a finite transverse width~\cite{Luscher:1980iy}, whose properties can be studied analytically within EST and tested numerically in lattice simulations.

The simplest EST model for the flux tube is the well known Nambu-Got\=o string~\cite{Nambu:1978bd,Goto:1971ce}, which predicts that at low temperatures the squared width of the flux tube grows logarithmically with the interquark distance~\cite{Luscher:1980iy}. This behaviour was observed in lattice simulations of various pure gauge theories~\cite{Caselle:1995fh, Zach:1997yz, Koma:2003gi, Gliozzi:2010zv, Bakry:2010zt}. In the proximity of the deconfinement transition (i.e., at higher temperatures but still in the confining regime) the same framework predicts a linear dependence on the interquark distance, with a slope diverging as the system approaches deconfinement~\cite{Allais:2009uos, Caselle:2010zs}; this analytical prediction also agrees with the results of lattice simulations~\cite{Caselle:2010zs, Gliozzi:2010jh}.

However, it is well established that the Nambu–Got\=o action is not the complete description of the EST. For example, universality theorems~\cite{Luscher:2004ib, Dubovsky:2012sh, Aharony:2013ipa} show that the first few terms in the large-distance expansion of any consistent effective string action coincide with those of the Nambu-Got\=o model, thus explaining its success at long distances. However at higher orders non-universal corrections appear, revealing the limits of this approximate description(see, e.g., Ref.~\cite{Caselle:2024zoh} for how such corrections can be studied on the lattice).

Another source of deviations from the Nambu-Got\=o model that is particularly interesting to study is due to the coupling between the massless string fluctuations and the massive, non-stringy (“intrinsic”) excitations of the underlying gauge theory. Exactly for this reason such corrections are non-universal, i.e., they depend on the particular gauge theory under study (see for instance the discussion in Refs.~\cite{Aharony:2024ctf,Caselle:2025vhx} for the three-dimensional $\mathrm{U}(1)$ case). It turns out that one of the major effects of these massive modes is the creation of an ``intrinsic'' width of the flux tube, distinct from the width generated by quantum fluctuations which is expected to be related to the inverse of the glueball mass. The goal is to use this intrinsic width, its behaviour as a function of temperature and its relation with the EST derived Gaussian shape of the flux tube as a tool to understand the intrinsic degrees of freedom which drive the formation of the confining flux tube in non-abelian gauge theories which, as it is well known, is a major open problem in this field. This program works perfectly in the simpler case of the $\mathrm{U}(1)$ model in three spacetime dimensions, where the confining mechanism is well known~\cite{Polyakov1977,GopfMack}, its effect on the intrinsic width can be evaluated exactly~\cite{Aharony:2024ctf} and is in good agreement with the results of numerical simulations~\cite{Caselle:2016mqu}.

One of the goals of the present paper is to try a similar approach for non-abelian gauge models in which the confining mechanism is unknown, by extracting precise estimates of the intrinsic width and then showing how they can be used to test possible confining mechanisms. We perform this analysis in the $(2+1)$-dimensional $\SU(2)$ pure gauge theory for different combinations of temperatures and lattice spacings. This model is an ideal testing ground for such studies: it shares the same infrared physics of the more complex four-dimensional non-abelian gauge theories while allowing for highly precise numerical determinations of the flux tube profile with reduced computational cost.

In particular the goal of this paper is to test the aforementioned expectation that the intrinsic width should be related to the inverse of the glueball mass, to investigate how it behaves as a function of the temperature and to study how it is combined with the Gaussian shape predicted by the EST.

A crucial step of our analysis is the precise determination of the intrinsic width, which is a challenging task from a numerical point of view. It turns out that the most sensitive approach (which we follow in this paper) relies on studying the transverse shape of the flux tube profile~\cite{Cardoso:2012aj, Cardoso:2013lla, Baker:2018mhw, Baker:2023dnn}. In the Nambu–Got\=o scenario, this profile is predicted to be Gaussian: consequently, any deviation from this kind of shape provides a clear signal of the intrinsic structure of the flux tube and offers a way to estimate its intrinsic width.

For reasons that will be clear later we study separately the two limits of low and high temperatures. In the low temperature limit we first compare our results with a description of the flux tube profile that is as model-independent as possible and extract in this way a conservative estimate for the intrinsic width. We then turn to a selection of confinement models proposed in the literature. We find that none of them provides a fully satisfactory, self-consistent description of the values extracted from our simulations. Our hope is that, by studying how these models deviate from the data, one may gain further insight into how they can be refined to describe the confining flux tube in this regime.

Then we perform the same comparison in the high-temperature regime, where the theoretical framework is quite different, since an exact solution can be obtained using dimensional reduction. Indeed, in this setup we find that our results not only are in perfect agreement with the model predictions, but that we are also able to perform a non-trivial self-consistency check with completely independent results obtained studying the interquark potential.

This paper is organized as follows: Section~\ref{sec:lattice_setup} is devoted to a discussion of our lattice gauge theory setup. In Section~\ref{sec:fluxtube_profile} we briefly discuss a few known results on the intrinsic width: first its analysis in the three-dimensional $\mathrm{U}(1)$ case, then its relationship with the dual superconductor model of confinement and finally the connection at high temperature with the behaviour of three point functions of spin models in two dimensions. In Section~\ref{sec:results} we present our results and compare them with various existing models and predictions. Finally we devote Section~\ref{sec:conclusions} to some concluding remarks and discussion of open issues. A preliminary account of our results was reported in Ref.~\cite{Verzichelli:2025cqc}.

\section{Lattice Setup}
\label{sec:lattice_setup}

We regularize the $\SU(2)$ Yang-Mills theory on a three-dimensional cubic lattice with a total volume $V = {L_s}^2 \times L_t$ and a lattice spacing $a$. We impose periodic boundary conditions along all directions with period $L_s = a \, N_s$ for the two space directions and $L_t = a \, N_t$ for the Euclidean time one. To each link of the lattice, we associate a group-valued variable $U_\mu(x)$, identified by the space-time coordinate $x$ and the direction $\mu$ of the link.

We adopt the Wilson discretization of the gluonic action:
\begin{equation}
    S[U] = \beta \sum_{x, \, \mu < \nu} \left( 1 - \real \Pi_{\mu \nu} \right),
\end{equation}
where $\beta$ is related to the inverse bare gauge coupling as $\beta = 4 / (a g^2)$, while $\Pi_{\mu\nu}$ is the plaquette operator, obtained by multiplying the four link variables along a minimal close path on the lattice and tracing over colour indices: 
\begin{equation}
    \Pi_{\mu \nu} = \tfrac{1}{2} \, \Tr [U_\mu(x) U_\nu(x+\hat{\mu}) {U_\mu}^\dagger(x+\hat{\nu}) {U_\nu}^\dagger(x)].
\end{equation}

It is well known that the three-dimensional $\SU(2)$ pure gauge theory exhibits a second order deconfinement transition at a non-zero physical temperature, which on the lattice is identified with the inverse size of the Euclidean time extent $T = 1 / L_t$. If we fix the number of sites $N_t$, we can finely tune the temperature of the system by varying the spacing $a$: in other words, by using a precisely known scale setting, we can relate the temperature $T$ directly with the bare coupling and thus with $\beta$. 
In particular, for a given value of $N_t$ the critical temperature $T_c$ is very precisely known from Ref.~\cite{Edwards:2009qw} by tuning the inverse coupling in the following way:
\begin{equation}
    \beta_c(N_t) = 1.5028(21) N_t + 0.705(21) - 0.718(49) \frac{1}{N_t}.
\end{equation}
Our discussion in this work will be strictly limited to the confined phase $T < T_c$: in particular, we will explore both the regime of small temperatures $T \ll T_c$, as well as the one close to the deconfinement transition $T \lesssim T_c$.

The $T < T_c$ phase is characterized by a vanishing expectation value of the Polyakov loop $P$, defined as the trace of the ordered product of all the time-like ($\mu = \hat{0}$) link variables at a given space point $\Vec{x}$:
\begin{equation}
    P = \frac{1}{2} \Tr \prod_{t = 1}^{N_t} U_0 \left(\vec{x}, t \right).
\end{equation}
The primary observable related to the properties of the flux tube between two static sources is the two-point correlator of Polyakov loops:
\begin{equation}
    G(R) = \left< \frac{1}{{N_s}^2} \, \sum_{\vec{x}} P \left(\vec{x}\right) \, P^\dagger \left(\vec{x} + R \hat i\right) \right>.
    \label{eq:two_pts}
\end{equation}
The choice of the space direction $\hat i = \hat 1, \hat 2$ is immaterial and it is possible to average over the two options. In what follows, we will always intend the two Polyakov loops to be separated along the direction $\hat i = \hat 1$, which we will refer to as the \emph{longitudinal} direction, while the perpendicular one ($\hat 2$), will be called the \emph{transversal} direction.

\begin{figure}
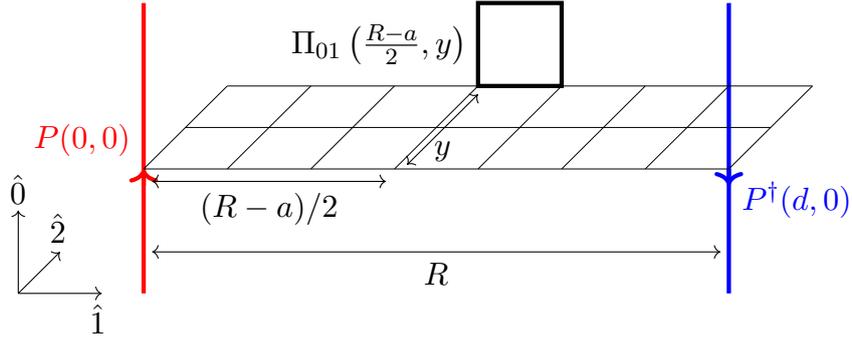

    \centering
    \include{poly_poly_plaq}
    \caption{Schematic representation of the three-point function $F_{\mu \nu}$ in Eq.~\eqref{eq:three_pts}, with $\hat \mu = \hat 0$ and $\hat \nu = \hat 1$. Thick lines indicate the traced Wilson lines: the two Polyakov loops (red, blue) and the plaquette operator (black).}
    \label{fig:poly_poly_plaq}
\end{figure}

We are interested in a lattice observable whose continuum limit corresponds to the square of the field strength tensor in the presence of a quark-anti-quark pair~\cite{Fukugita:1983du, DiGiacomo:1990hc}: in this work we use
\begin{equation}
    F_{\mu \nu}(R, y) = \left< \frac{1}{{N_s}^2} \, \sum_{\vec{x}} P \! \left(\vec{x}\right) \, \Pi_{\mu\nu} \! \left(\vec{x} + \vec{l} \right) \, P^\dagger \! \left(\vec{x} + R \hat 1\right) \right>,
\label{eq:three_pts}
\end{equation}
where $\vec{l}$ is the vector $(R - a) / 2  \,  \hat 1 + y \, \hat 2$. We will always choose $R$ to be odd in units of the lattice spacing $a$, so that $\vec{l}$ is a vector on the lattice: this choice allows the plaquette operator appearing in Eq.~\eqref{eq:three_pts} to be equally distant from the two Polyakov loops. A schematic illustration of $F_{\mu \nu}$ can be found in Fig.~\ref{fig:poly_poly_plaq}.
Let us stress that to avoid systematic effects we did not connect (see Ref.~\cite{DiGiacomo:1990hc} for comparison) the plaquette with the Polyakov loop with a chain of gauge links. This choice ensures the final observable we define below correctly reproduces the trace of the square of the component of the field-strength tensor that corresponds to the $\mu$ and $\nu$ indices in Eq.~\eqref{eq:three_pts} (which is not obvious for other possible choices). Furthermore we did not perform any smearing or smoothing of the gauge configuration before the measurements. A discussion on the effects of these procedure on the measure of the observables we are interested in can be found in Ref.~\cite{Battelli:2019lkz}.

To compute the three-point function of Eq.~\eqref{eq:three_pts} numerically, we fix $\hat \mu = \hat 0$ and $\hat \nu = \hat 1$: this choice corresponds to probing the chromo-electric field and, in particular, its longitudinal component. As studied in Ref.~\cite{Bonati:2020orj}, this setup provides the best signal, so we will use it for the rest of this work.
Furthermore, in order to obtain an observable that vanishes in the $y \to \infty$ limit, we normalize the three-point function $F_{01}$ by the two-point function of Eq.~\eqref{eq:two_pts} and subtract the expectation value of the plaquette:
\begin{equation}
\label{eq:profile}
    \rho (R, y) = \frac{F_{01}(R, y)}{G(R)} - \left< \Pi_{01} \right>.
\end{equation}
The orientation of the plaquette operator is not relevant, as long as it is time-like, i.e., one of the two directions is the temporal one.

After fixing the distance $R$ between the two Polyakov loops, we call $\rho$, now a function of the transversal displacement $y$, the \emph{profile} of the flux tube. In this work we performed numerical Monte Carlo simulations of the theory and determined the profile $\rho$ in a variety of settings.  In particular, we explored different values of the temperature $T$, of the lattice spacing $a$ and of the distance $R$ between the static sources.

These values of $\rho(R,y)$ are the data that we fit in Section~\ref{sec:results} with the models that we  discuss in the next section. In particular we are interested in modelling the deviations with respect to a purely Gaussian behaviour of $\rho$ as a function of the transverse direction $y$. These deviations manifest themselves as exponentially decreasing tails for large values of $y$ and allow us to define a new physical scale, that we denote in the following as $\lambda$. This is exactly the intrinsic width of the flux tube in which we are interested in the present paper. It should be noticed that, from a lattice gauge theory perspective, the correlator of two Polyakov loops will have some overlap with the destruction operator of a scalar glueball sourced by the plaquette. This leads to the expectation of finding a decaying length $\lambda$ similar to the inverse mass of the lightest glueball $1 / M_0$, which is also suggested by some analogies that we present in Sections~\ref{sec:profile_u1} and \ref{sec:Hol}. Let us, however anticipate that the main decaying length we extract numerically will be larger than such an inverse mass ($\lambda \, M_0 > 1$), which of course does not exclude the presence of a faster decaying mode.

\section{A few results on the profile of the flux tube and its intrinsic width}
\label{sec:fluxtube_profile}

In this section we discuss a few theoretical results on the profile of the flux tube and its intrinsic width which we will then use in the next sections to fit our data. We concentrate in particular in the case of $(2+1)$-dimensional gauge models, in which the EST is characterized by only one transverse degree of freedom.
We first review the Nambu-Got\=o predictions (Section~\ref{sec:ngwidth}) and how they are modified by higher order terms in the EST action. Then we discuss the behaviour of the intrinsic width in the $\mathrm{U}(1)$ gauge model in $(2+1)$ dimensions (Section~\ref{sec:profile_u1}), in the dual superconductor scenario of confinement (Section~\ref{sec:clem}), and in the high temperature regime of the $\SU(2)$ gauge model in $(2+1)$ dimensions (Section~\ref{sec:sy}); finally, Section~\ref{sec:Hol} is devoted to a discussion of some holographic results on the intrinsic width.

\subsection{The Nambu-Got\=o prediction for the width of the flux tube}
\label{sec:ngwidth}

\subsubsection{Low temperature behaviour}
\label{sec:lowT}

The Nambu-Got\=o EST is simple enough to allow some exact calculation in the large $\sqrt\sigma R$ limit, where, as defined above, $R$ is the interquark distance and $\sigma$ the string tension.  In this limit the Nambu-Got\=o model flows to the exactly solvable theory of a 2-dimensional massless free boson. Indeed, it is possible to show that at low temperatures (which means $T \ll T_c$ where $T_c$ is the deconfinement temperature in the gauge theory) the profile of the string, which is fully due to quantum fluctuations, is a pure Gaussian with a width $w$ whose square increases logarithmically with the distance~\cite{Luscher:1980iy}:
\begin{equation}
    \label{eq:EST_logbroad}
    w^2 = \frac{1}{2 \pi \, \sigma} \, \log \! \left( \frac{R}{R_0} \right)
\end{equation}
where $R_0$ is a non-universal scale which emerges from the regularization of the two point function which is used to evaluate the width.
This scale will play an important role in the following since it can be related (in a rather non obvious way) to the intrinsic width which we are studying. 

This logarithmic increase has been confirmed over the years by several studies~\cite{Caselle:1995fh, Zach:1997yz, Koma:2003gi, Gliozzi:2010zv, Bakry:2010zt} and precise numerical estimates were obtained for the value of $R_0$. Importantly, this quantity was shown to have a good scaling behaviour in the continuum limit and thus it is expected to represent a new physical scale of the model. 
These same studies however also revealed deviations from the Gaussian profile, which we now understand as due to the intrinsic width of the flux tube (which we discuss later) and at the time were typically fitted with phenomenological expressions.

Recently, there have been several efforts to go beyond the free boson limit, leading only to an exact calculation of the next-to-leading order in $1/R$~\cite{Gliozzi:2010zv} but with no results for the higher order terms. Similarly, no analytical prediction exists for the effect on the string width of the higher order corrections beyond the Nambu-Got\=o action, which are known to characterize the EST in non-abelian gauge theories~\cite{Caristo:2021tbk,Caselle:2024zoh}.

\subsubsection{High temperature behaviour}

In the high temperature regime (which means $T$ slightly below $T_c$, i.e, still in the confining phase) the free bosonic action is known to show a linear broadening~\cite{Allais:2009uos, Caselle:2010zs}
\begin{equation}
    \label{eq:EST_linbroad}
    w^2 = \frac{R}{4 \,L_t \sigma} + C
\end{equation}
where $C$ is a non-universal constant. 
In this regime, besides an exact next-to-leading order calculation~\cite{Gliozzi:2010jh}, there is also a conjecture for the behaviour at all orders of the Nambu-Got\=o width~\cite{Caselle:2010zs} (supported by the Svetitsky--Yaffe approach discussed in Section~\ref{sec:sy}) which simply amounts to substitute the string tension in Eq.~\eqref{eq:EST_linbroad} with the one containing the finite temperature corrections
\begin{equation}
\label{eq:conjecture}
\sigma(T)=\sigma\sqrt{1-\frac{T^2}{T_c^2}}
\end{equation}
which was recently confirmed~\cite{Caselle:2023mvh,Caselle:2024ent} by a set of flow-based simulations directly in the Nambu-Got\=o model.  

\subsubsection{Effective string numerical results from flow-based simulations}
\label{sec:profile_snf}

A drastic improvement in this context has been recently achieved by directly simulating the Nambu-Got\=o action (and its higher order corrections) on the lattice~\cite{Caselle:2023mvh,Caselle:2024ent}\footnote{A more detailed discussion of these simulations is reported in Appendix~\ref{app:SNF}.} using Normalizing Flows methods~\cite{rezende2015variational,Albergo:2019eim,Nicoli:2020njz,Cranmer:2023xbe}.
The behaviour observed in the free bosonic limit is confirmed also for the whole Nambu-Got\=o action. In particular:
\begin{itemize}
\item
The shape of the flux tube is perfectly Gaussian.
\item
The scale $R_0$ is of the order of the lattice spacing and does not grow when approaching the continuum limit.
\end{itemize}
Moreover in Ref.~\cite{Caselle:2024ent} it was shown that the same behaviour is present also when higher order terms (in particular a term proportional to the fourth power of the extrinsic curvature -- the first one allowed by Lorentz invariance) are added to the Nambu--Got\=o action.
This suggests that these terms are not sufficient to explain the observed deviation with respect of the Gaussian shape (i.e., the intrinsic width) and that the interaction of the EST degrees of freedom with the bulk, massive degrees of freedom of the gauge theory must be addressed in all its complexity. An interesting example of this approach is represented by the $\mathrm{U}(1)$ model in $(2+1)$ dimensions discussed in detail in the next subsection.

\subsection{The string width in the \texorpdfstring{$\mathrm{U}(1)$}{U(1)} lattice gauge theory in three dimensions}
\label{sec:profile_u1}

In Ref.~\cite{Aharony:2024ctf}, the profile of the flux tube for the three dimensional  $\mathrm{U}(1)$ gauge theory was obtained analytically leveraging the results from Ref.~\cite{Polyakov:1975rs} on the classical profile of the flux tube. The leading quantum correction is given by Gaussian fluctuations of the center of this classical profile: the result, at this order of approximation, is the convolution between a Gaussian and the classical profile.

The classical term is exactly computed as the profile of an interface given by condensed monopoles. The solitonic solution is
\begin{equation}
    F_{01}^{\text{(cl)}} =
    \frac{\mathrm{e}^2}{\pi \, \lambda} \,
    \mathrm{sech} \! \left( \frac{y}{\lambda} \right)
\end{equation}
where $\mathrm{e}$ is the QED coupling ($\mathcal{L} = -\tfrac{1}{4} F_{\mu\nu}^2 / \mathrm{e^2}$); it features a characteristic length $\lambda$ (that we identify with the intrinsic width) which is the inverse of the mass of the particle in the bulk (photon/``photoball''), i.e., the mass gap. 

In our notation, the result from Ref.~\cite{Aharony:2024ctf} can be written as:
\begin{equation}
    \braket{F_{01}} = \frac{1}{\sqrt{2 \, \pi^3}} \, 
    \frac{\mathrm{e}^2}{\lambda \, s} \int \diff t 
    \, e^{-\tfrac{1}{2} t^2 / s^2} \,
    \mathrm{sech} \! \left( \frac{y - t}{\lambda} \right),
\label{U1}
\end{equation}
where $s^2 \propto \log(M \, R)$, for some cutoff $M$, is the square width of the Gaussian fluctuation\footnote{As usual the charges are separated along the $\hat 1$ direction by a (large) distance $R$ and the field is measured midway between the charges with a transverse (along $\hat 2$) displacement $y$.}.
In this case the intrinsic width is by construction the inverse of the ``photoball'' mass. The non-trivial aspect of the analysis is in the way it is combined with the Gaussian fluctuations of the confining flux tube, i.e., in the particular form of Eq.~\eqref{U1} above.

\subsection{The dual superconductor scenario of confinement and the Clem ansatz}
\label{sec:clem}

In the past years the proposal by Polyakov of a confinement mechanism driven by monopole condensation was extended also to non-abelian gauge theories and led to the so called ``dual superconductor model'' of confinement~\cite{Mandelstam:1974pi, tHooft:1981bkw}. 
In this context an expression for the profile of the flux tube was proposed in Refs.~\cite{Cea:2012qw, Cea:2017ocq}, based on a phenomenological model proposed long ago by Clem~\cite{Clem:1975ohd} to describe the profile of a vortex in a superconductor. The explicit form of the Clem formula for the flux tube profile is, in our notation:
\begin{equation}
\label{eq:clem}
    \rho(y) = A^{\rm (Clem)} \, K_0 \! \left( \frac{\sqrt{y^2 + \xi^2}}{\lambda} \right),
\end{equation}
where $K_0$ is the modified Bessel function of the second kind, of order zero. 
Besides the multiplicative normalization constant $A^{\rm(Clem)}$, two independent parameters appear: $\xi$ and $\lambda$. In the dual superconductor model for confinement, they correspond to two typical lengths of a vortex: $\xi$ is the radius of the core, while $\lambda$ is the London penetration length. 
It is easy to see that Eq.~\eqref{eq:clem} describes a Gaussian shape for $y \ll \xi$ smoothly connected to the exponentially decaying tail for $y \gg \xi$ and can be considered as the analogous in this context of Eq.~\eqref{U1} for the abelian case. In this case $\lambda$ cannot be directly derived from the underlying gauge theory, since the dual superconductor model is only an effective description of the confining phase of the theory. Even if it is closely related to the interaction of the string with degrees of freedom in the bulk and their mass, we do not expect an exact coincidence as in the $\mathrm{U}(1)$ case.

Note, to avoid confusion, that the length $\xi$ is not the coherence length of the superconductor (usually denoted by the same letter), that in our notation can be expressed instead as $\kappa \, \lambda$, with $\kappa$ being the Ginzburg--Landau parameter. The importance of the latter in the dual superconductor picture is due to the radically different behaviour of a superconductor with $\kappa < 1 / \sqrt{2}$ (type I) from one with $\kappa > 1 / \sqrt{2}$ (type II). The Ginzburg--Landau parameter $\kappa$ can be computed with the formula reported in Ref.~\cite{Clem:1975ohd} as
\begin{equation}
    \label{eq:clem_ginsland}
    \kappa = \frac{\sqrt{2} \, \lambda}{\xi} \, \sqrt{1 - {K_0}^2 \! \left( \frac{\xi}{\lambda} \right) \Big/ {K_1}^2 \! \left( \frac{\xi}{\lambda} \right)}.
\end{equation}

In the past a controversy arose on the best value of $\kappa$ for describing a superconductor dual to the vacuum of the $(3+1)$ dimensional $\SU(3)$ Yang-Mills theory, with no general agreement on the type of the superconductor (see for example Refs.~\cite{Cea:2012qw, Cardoso:2010kw}). In the original model $\kappa$ has no dependence on the length of the flux tube $R$; as we will see, for the $(2+1)$-dimensional $\SU(2)$ case, when we try to extract the value of $\kappa$ from the flux tube profile, it appears to be a decreasing function of $R$ tending to zero in the infinite length limit.

\subsection{The high temperature behaviour of the intrinsic width and the Svetitsky--Yaffe mapping}
\label{sec:sy}

In the high temperature limit (i.e., near the deconfinement transition yet still in the confining phase) an exact prediction for the shape of the flux tube can be obtained thanks to the fact that the deconfinement transition of the $(2+1)$-dimensional $\SU(2)$ model is of the second order and is in the same universality class of the magnetization transition of the 2-dimensional Ising model. In the neighbourhood of the critical point the two- and three-point functions that we are interested in are mapped into the spin-spin and spin-spin-energy correlators of the 2-dimensional Ising model. 
This is known as the Svetitsky--Yaffe (SY) mapping~\cite{Svetitsky:1982gs} and it is a powerful tool to describe the physics in the vicinity of the deconfinement transition. The spin-spin and spin-spin-energy correlators for the Ising model and their application to the study of the shape of the flux tube was discussed in Refs.~\cite{Caselle:2006wr, Caselle:2012rp}. 

It turns out that the two- and three-point functions can be expressed in terms of only one length scale $\lambda$.
The spin-spin correlator is
\begin{equation}
    \frac{ \left( F_1^\sigma \right) ^2}{\pi} \,
    K_0 \! \left( \frac{R}{2\lambda} \right),
    \label{eq:Ising_2pts}
\end{equation}
while the spin-spin-energy correlator is:
\begin{equation}
    \frac{ 2 \, \left( F_1^\sigma \right)^2 \, R}{4 \, y^2 + R^2}
    \, \exp \! \left( -\frac{\sqrt{y^2 + (R / 2)^2}}{\lambda} \right)
    \label{eq:Ising_3pts}
\end{equation}
where $K_0$ is, as above, the modified Bessel function of order zero, and $F_1^\sigma$ is a form factor which can be evaluated exactly, but does not play any role in our analysis since it simplifies in the ratio. The final answer for the shape $\rho$ of the flux tube close to the deconfinement transition is
\begin{equation}
    \rho(d, y) = A^{\rm(SY)} \, \frac{2 \pi R}{4 l^2} \, \frac{\exp(-l / \lambda)}{K_0\{R / (2 \, \lambda) \}},
    \label{eq:SY_profile}
\end{equation}
where $l$ is the distance between each of the Polyakov loops and the plaquette. Since the plaquette is one lattice spacing wide, in our fit we will use the following expression for $l$:
\begin{equation}
    l = \sqrt{\left(\frac{R - a}{2}\right)^2 + y^2}.
\end{equation}

Notice that in this case $\lambda$ is not directly related to the inverse of the glueball mass but instead to the inverse of the ground state energy $E_0$ of the string. The latter in turn is related to the finite temperature string tension $\sigma(T)$ defined in Eq.~(\ref{eq:conjecture}) through the relation $E_0=\sigma(T)/T$; we note that in this regime this quantity is a much lower energy scale than the glueball mass.

Notice that the only free parameters of Eq.~(\ref{eq:SY_profile}) are $A^{\rm (SY)}$ and $\lambda$.
Moreover, we expect that they do not depend on $R$, a prediction which represents a strong consistency test of the whole expression.

It is interesting to compare this high temperature result with the Clem formula. The two expressions are similar, but there are two important differences:
\begin{itemize}
\item
The Svetitsky--Yaffe mapping has one free parameter less compared to the Clem formula, since the role played by the variational length $\xi$ in the Clem formula is played here by $R$, the distance between the two Polyakov loops.
\item
Both expressions are characterised at large distances by an exponential decrease divided by a power law. In the case of the Clem model, though, this power is $\sqrt{y}$, while in the SY mapping, the subleading power is the square of the argument of the exponential (thus proportional to $l^2 \sim y^2$ for large $y$):
\begin{align}
    \rho^\text{(Clem)} &\propto \frac{\exp(y / \lambda)}{\sqrt{y}} \, \left( 1 + \mathcal{O}\left(\frac{1}{y} \right) \right),\\ 
    \rho^\text{(SY)} &\propto \frac{\exp(y / \lambda)}{y^2} \, \left( 1 + \mathcal{O}\left(\frac{1}{y^2} \right) \right).
\end{align}
\end{itemize}

\subsection{Holographic predictions for the flux tube shape}
\label{sec:Hol}

A different approach to the problem is represented by holographic duality. 
In a recent paper~\cite{Canneti:2025afi} a holographic description of the flux tube was obtained, relying on the duality between large-$N$ Yang--Mills and a gravity background on Anti-de Sitter space. This work extends previous results of Ref.~\cite{Danielsson:1998wt}. The main outcome of Ref.~\cite{Canneti:2025afi} is a refined estimate of the intrinsic width which is shown, via a holographic mapping, to coincide with the inverse of the mass of the lowest lying glueball state. We shall see below that, despite the fact that we are very far form the large $N$ limit, our numerical estimates for the intrinsic width in the $\SU(2)$ theory are indeed very similar to the inverse of the lowest glueball mass of the model.

\section{Numerical results}
\label{sec:results}

The simulation details for all the results we obtained are displayed in Tab.~\ref{tab:DetailsLowT} and \ref{tab:DetailsHighT}. The reported values of the lattice spacing in units of the string tension are obtained interpolating the data in Ref.~\cite{Athenodorou:2016ebg}.

\begin{table}[ht]
    \centering
    \begin{tabular}{|c|c|c|c|c|c|}
    
\hline
$\beta$   & $a \, \sqrt{\sigma}$ & $N_t$ & $T / T_c$ & $N_s$ & $R / a$           \\ \hline
8.768     & 0.16702(28)          & 24    & 0.23      & 80    & 9, 11, 13, 15 \\
\hline
\multirow{3}{*}{10.865} & \multirow{3}{*}{0.13137(86)}
                                 & 20    & 0.34      & 96    & 9, 11, 15, 19 \\
          &                      & 30    & 0.23      & 96    & 9, 11, 13, 15 \\
          &                      & 60    & 0.11      & 60    & 9             \\
\hline
12.9625   & 0.10870(62)          & 36    & 0.23      & 120   & 11, 13, 15, 17 \\
\hline

    \end{tabular}
    \caption{Simulation details for the simulations at the three smallest values of the temperature we considered. For these simulations we used the L\"uscher-Weisz ``multilevel'' algorithm~\cite{Luscher:2001up} with $\mathcal{O}(10^4)$ updates in an internal level and $\mathcal{O}(100)$ updates in the external one.}
    \label{tab:DetailsLowT}
\end{table}

\begin{table}[ht]
    \centering
    \begin{tabular}{|c|c|c|c|c|rl|}
    \hline
    $\beta$ & $a \, \sqrt{\sigma}$ & $N_t$ & $T / T_c$ & $N_s$ &
    \multicolumn{2}{c|}{$R / a$} \\ \hline
    \multirow{3}{*}{10.865} & \multirow{3}{*}{0.13137(86)}
                         & 14 & 0.49 & 60  &  7 & 21 \\
           &             & 10 & 0.68 & 96  & 11 & 21 \\
           &             &  8 & 0.85 & 96  &  9 & 21 \\ \hline
    11.914 & 0.11881(81) & 11 & 0.68 & 96  & 11 & 21 \\ \hline 
    12.962 & 0.10870(62) & 12 & 0.68 & 96  & 11 & 17 \\ \hline
    13.424 & 0.10486(51) & 10 & 0.85 & 120 &  9 & 21 \\ \hline
    14.011 & 0.10046(35) & 13 & 0.68 & 120 &  9 & 21 \\ \hline
    \end{tabular}
    \caption{Simulation details at the highest values of the temperature we considered. The profile was computed at all the odd values of $R / a$ between the indicated extrema, including them.}
    \label{tab:DetailsHighT}
\end{table}

For all the values of $\beta$, $T/T_c$ and $R$ that we studied we observed strong deviations from the Gaussian shape predicted by the naive EST model, compatible with the presence of an intrinsic width in the confining flux tube. 
Due to the different behaviour of the flux tube shape in the low- and high-temperature regimes we discuss them separately.
We devote the next subsection to quantify and characterize these deviations and to a model independent evaluation of the intrinsic width. Then, in the following subsections, we compare our results with the models discussed in Section~\ref{sec:fluxtube_profile} above.

We are attaching to this manuscript the numerical results for $\rho(R,y)$ for all the values of $\beta$ and $T/T_c$ that we studied, as well as a script to read them and reproduce some of our results\footnote{In order to correctly reproduce some of our fits, additional information about the correlation of the data is needed; we will be happy to provide them on request.}.

\subsection{Flux tube profile in the low temperature regime}

\begin{figure}
    \centering
    \includegraphics[width=\textwidth]{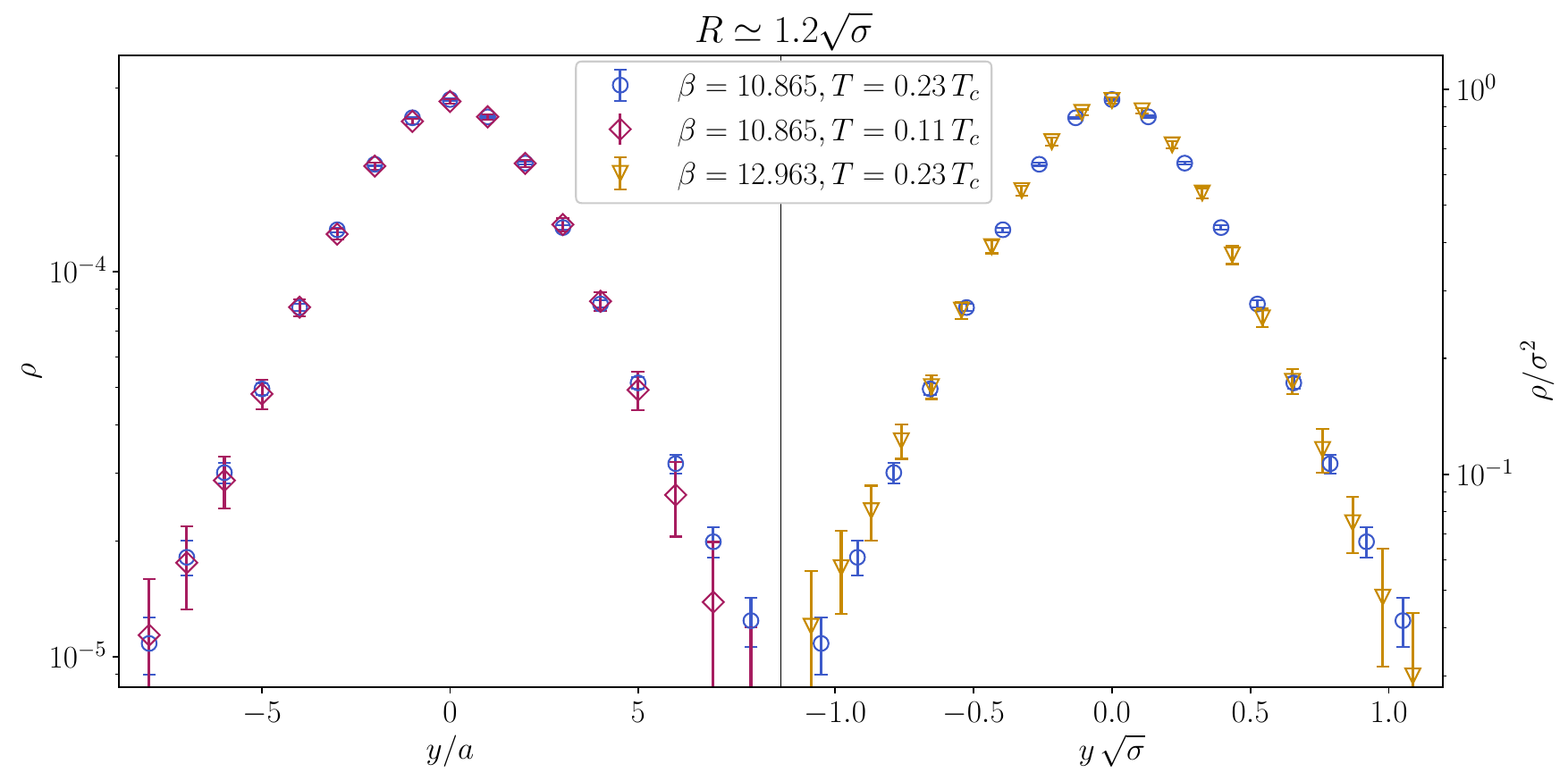}
    \caption{The profile of the flux tube for fixed lattice spacing and different temperature $T$ (left panel) and vice versa (right panel). Note that, in order to compare between different values of the lattice spacing, the profile has been rescaled with the square of the string tension. The profiles obtained with $\beta = 10.865$ (left panel) correspond to $R = 9 \, a = 1.18 / \sqrt{\sigma}$, while for the one at $\beta=12.963$ the distance was set to $R = 11 \, a = 1.20 / \sqrt{\sigma}$.
    }
    \label{fig:compare_profile}
\end{figure}

We start by analysing the data for the flux tube profile defined in Eq.~\eqref{eq:profile} obtained from simulations at low temperature; we report some representative results in Fig.~\ref{fig:compare_profile}. We can immediately appreciate the stability of the results when changing the temperature from $0.23 \, T_c$ to $0.11 \, T_c$ or the lattice spacing from $a \sqrt{\sigma} = 0.13$ to $0.11$.
The main feature of all the numerical results for $\rho$ (including those in Fig.~\ref{fig:compare_profile}) is that for all the values of temperature and lattice spacing that we explored, we found, as anticipated, significant deviations from a Gaussian profile. Such deviations show up as exponentially decreasing tails for large values of $y$ and appear as straight lines in the log scale of Fig.~\ref{fig:compare_profile}. As mentioned above, these exponential tails allow us to define a new physical scale, which is exactly the intrinsic width $\lambda$. 

\begin{figure}[htb]
    \centering
    \includegraphics[width=0.58333\textwidth]{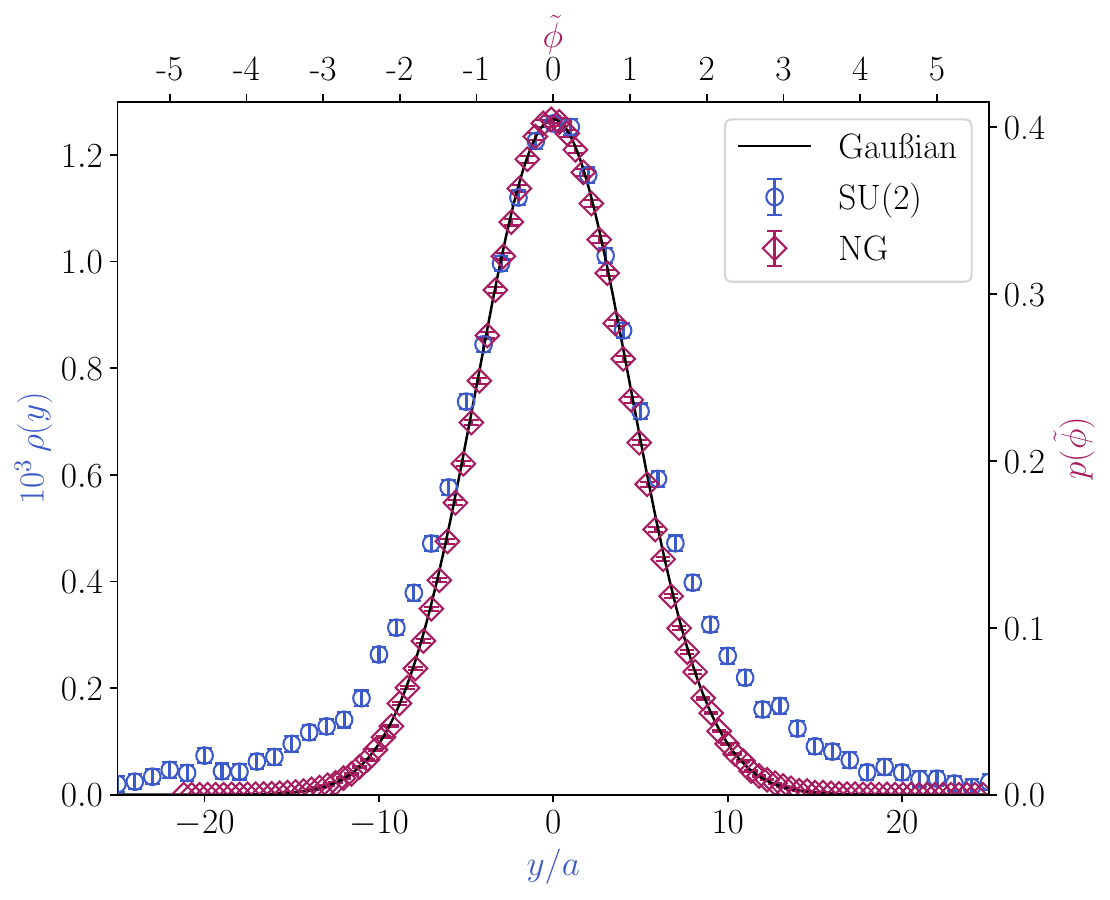}
    \caption{Comparison of the profile from a simulation of the $(2+1)$-dimensional $\SU(2)$ gauge theory (at $\beta = 10.865$) with one from direct simulation of the Nambu-Got=o EST. In both cases, the length of the string equals $9a$ and the temporal extent is $60a$.}
    \label{fig:SU2_vs_NG}
\end{figure}

This behaviour is radically different from what is expected from the theory of a fluctuating string and what was numerically observed in Ref.~\cite{Caselle:2024ent}, even in presence of corrections to the Nambu-Got\=o action. A comparison between the profile obtained in this work for the gauge theory, and that obtained from direct simulation of the string theory is represented in Fig.~\ref{fig:SU2_vs_NG}. The profiles were scaled so that their Gaussian peaks have the same heights and width. It is clear that the profile extracted from the gauge theory has a slower decay at large $y$. 

Our interpretation is the following: the shape of the flux tube that we observe is due to the interaction between the degrees of freedom of the effective string and the degrees of freedom of the gauge theory. Thus, we expect the length $\lambda$ of the exponential decay of the tails to be related (even if not to coincide, as it was the case for the abelian $\mathrm{U}(1)$ model in $(2+1)$ dimensions) to the inverse of the lowest glueball mass of the theory. Our goal is to use such features of the flux-tube profile to shed some light on the microscopic mechanism driving confinement in the gauge theory.

\subsection{A first estimate of the intrinsic width \texorpdfstring{$\lambda$}{λ}}
\label{sec:fit_exp}

In this section we aim not only to quantify these deviations, but also to characterize the shape of the profile itself. 
To this end, we estimated the ``moments'' of $\rho$, intended as a (non-normalized) probability distribution, in analogy with the methods discussed in Ref.~\cite{Caselle:2024ent}.
In particular we are interested in:
\begin{itemize}
    \item the integral $I$ of the profile
    \begin{equation}
        I = \int \diff y \, \rho(y),
    \end{equation}
    \item the second moment $w^2$, corresponding to the (square of) the total width of the flux tube
    \begin{equation}
        w^2 = \frac{1}{I} \int \diff y \, y^2 \, \rho(y),
    \end{equation}
    \item the fourth moment $\mu_4$, to quantify the deviations from a Gaussian
    \begin{equation}
        \mu_4 = \frac{1}{I} \int \diff y \, y^4 \, \rho(y).
    \end{equation}
\end{itemize}
From these quantities we compute the so-called Binder cumulant $b$, which can be used as a parametrization of non-Gaussianity:
\begin{equation}
\label{eq:binder_def}
    b = 1 - \frac{\mu_4}{3 \, (w^2)^2}.
\end{equation}

In order to realize this program, we need a way to extract from our data for the profile $\rho$ at discrete values of $y$ a function of a continuous variable that can be integrated: the goal in this subsection is to compute the moments with sufficient precision and with a limited set of assumptions. 

The simplest model-independent approach would be to interpolate the profile data $\rho(y)$ with a spline, for instance a cubic polynomial spline. However, this approach leads to a very noisy behaviour of the interpolating function at the tails of the distribution\footnote{If we were interested just in the integral $I$ of the profile, we could just neglect the tails of the profile, since they give very little contribution. However, this strategy is not viable to compute higher moments (in particular the fourth one), since they are quite sensitive to the tails.}. The reason is twofold: first and most important, the exponential decay of the tails of the profile cannot be captured by a polynomial interpolation; secondly, the tails contain the points with the largest relative error and thus any kind of interpolation in that region results to be very noisy.

For a meaningful analysis of the profile, it is necessary to make some assumptions on the behaviour of the tails. As anticipated, we will assume that for large enough transverse displacement $y$, the profile can be approximated as a decreasing exponential, i.e., $\rho (y) \sim \exp(-|y| / \lambda)$. Here we introduce for the first time the characteristic length $\lambda$, which we will later identify with the intrinsic width.

\begin{figure}[ht!]
    \centering
    \includegraphics[width=0.58333\textwidth]{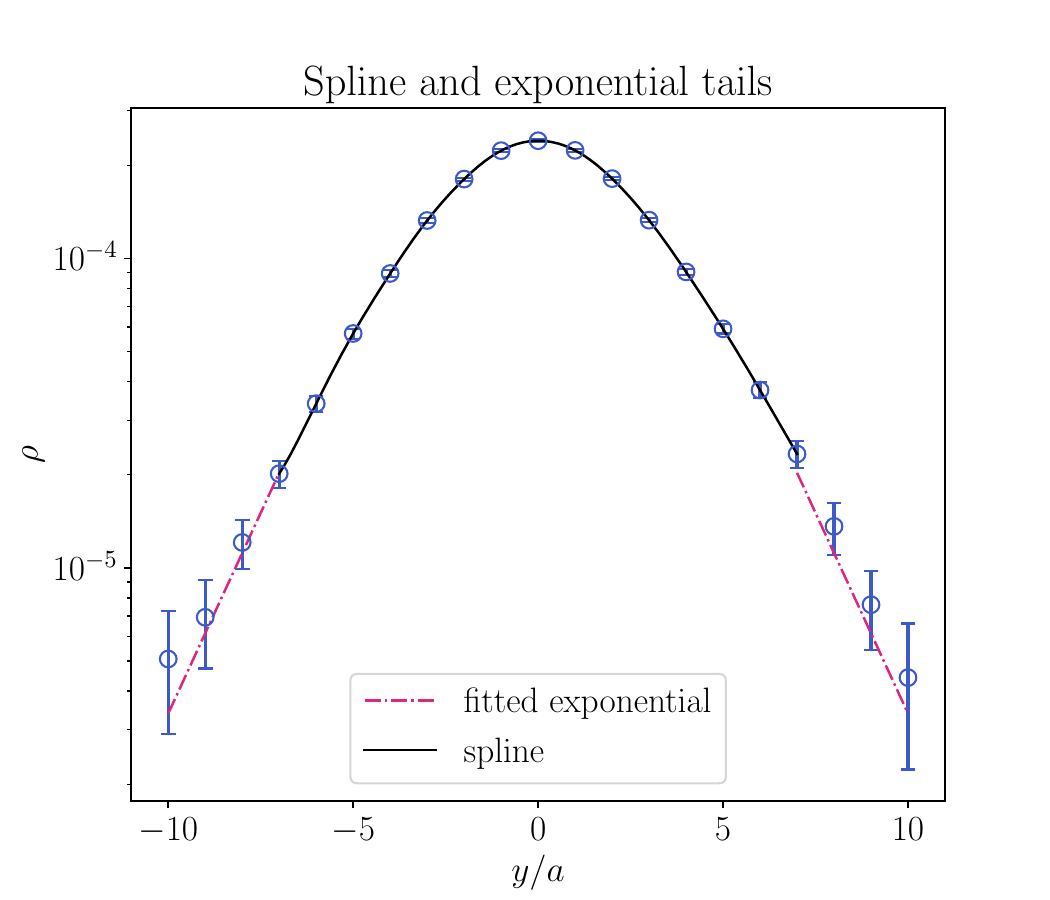}
    \caption{Example of spline interpolation of the Gaussian peak (solid black line) and exponential fit of the tails (red dash-dotted line). The profile was obtained from the simulation at $\beta = 10.865$ and $T = 0.23 \, T_c$, with $R = 11 a$.}
    \label{fig:exp_fit_example}
\end{figure}

The simplest way to make use of this assumption is to choose a value $y_\intr$, such that for $|y| > y_\intr$ we assume the exponential decay to be dominant over any other behaviour\footnote{The choice of $y_\intr$ is due to the interpretation of the exponential decay as the manifestation of the intrinsic width of the flux tube, since $|y| > y_\intr$ is the region where the effect of such intrinsic width are most evident.}. Following this assumption, we use for the numerical results for $\rho(y)$
\begin{itemize}
    \item a fit in $\rho(y) = A^\text{(exp)} \, \exp(-|y| / \lambda)$ for $|y| > y_\intr$
    \item a cubic spline interpolation for $|y| < y_\intr$.
\end{itemize}
More details on this procedure are given in Appendix~\ref{app:details_exp_fit}. In this case all the relevant integrals of the form $\int \diff y \, y^n \, \rho(y)$ can be analytically computed both in the tail region, where they reduce to $(- \diff / \diff m)^n \exp(-m \, x) / m$, and in the central interval, where they are piece-wise integrals of polynomials. We report some representative results for the $\lambda$ parameter, $w^2$ and $b$ in Tab.~\ref{tab:exp_fit_example}, while the full results for the fits are shown in Tab.~\ref{tab:exp_fit_details}; in Fig.~\ref{fig:exp_fit_example} we also show an example of this combination of fit and interpolation.

\begin{table}[ht!]
    \centering
    \textbf{Model independent method: results at $\beta = 10.865$} \\ [0.5em]
    \begin{tabular}{|c|c|c|c|c|c|}
    \hline
    $N_t$ & $R / a$ & $y_\intr / a$ & $\lambda / a$ & $w^2 / a^2$ & $b$ \\
    \hline
    60 &  9 & 5 & 1.92(26) & 9.78(19)   & -0.561(24) \\
    \hline
    \multirow{4}{*}{30} 
       &  9 & 6 & 1.84(10) & 9.577(43)  & -0.504(13) \\
       & 11 & 7 & 1.67(25) & 10.733(87) & -0.286(14) \\
       & 13 & 7 & 2.06(23) & 13.600(98) & -0.389(13) \\
       & 15 & 7 & 2.08(32) & 15.00(21)  & -0.318(13) \\
    \hline
    \multirow{4}{*}{20}
       &  9 & 6 & 1.79(11) & 9.174(54)  & -0.519(19) \\
       & 11 & 7 & 1.82(21) & 10.88(12)  & -0.394(30) \\
       & 15 & 8 & 1.75(29) & 13.69(26)  & -0.204(33) \\
       & 19 & 8 & 1.81(34) & 16.51(27)  & -0.135(22) \\
    \hline
    
    \end{tabular}
    \caption{Results from the analysis performed by fitting the tails with an exponential, described in Section~\ref{sec:fit_exp}, on the profiles we collected at $\beta = 10.865$. $\lambda$ does not show significant deviations, while the total width $w^2$ increases and Binder cumulant $b$ decreases with the distance $R$.}
    \label{tab:exp_fit_example}
\end{table}

Once the optimal value of $y_\intr$ is chosen, we may extract from the fits of the tail our best estimate for the intrinsic width, which we fully report in Tab.~\ref{tab:exp_fit_details}. The errors on $\lambda$ and values of the $\chi^2$ are in general rather large, possibly suggesting that our model is too simple to capture the physics completely. However, since these values were obtained with the minimal amount of assumptions, we consider them as a sort of ``reference value'' to be compared with other approaches.

First, we notice that the intrinsic widths of flux tubes with different lengths $R$ on the same lattice (i.e. at fixed $\beta$ and $T$) are in agreement with each other within at most two standard deviations. The lack of dependence on $R$ is a defining feature of the intrinsic width and means that only the central (Gaussian) part of the profile is sensitive to the length of the flux tube. 
At $\beta = 10.865$ we performed simulations at three values of the temperature: $T / T_c = 0.11, 0.23$ and $0.34$. The lack of a dependence of our results (particularly of $\lambda$) on the temperature suggests the absence of any thermal effects at this level of precision.

Finally, once it has been rescaled to some natural units, for example the string tension, the intrinsic width does not show a dependence on the lattice spacing (see Tab.~\ref{tab:ScaledLambdaExp}). This means that it is a well defined physical quantity in the continuum limit. We can extract a value taking a weighted average of the fit results at $T / T_c = 0.11, 0.23 $ and $0.34$ at the different lattice spacings, obtaining $\lambda\sqrt{\sigma} = 0.256(29)$. The estimation of the uncertainty is rather conservative and takes into account both the statistical error and the dispersion of the data points. In order to find more precise determination of $\lambda$, stronger assumptions on the fitting model are needed.

A very similar length has been recently found probing the flux tube with a completely different methodology~\cite{Amorosso:2024glf, Amorosso:2024leg}, in which it is possible to study the excess entanglement entropy of a region due to the presence of the flux tube. Its derivatives with respect to the extent of the considered region in the transverse direction ($\hat x$ in the notation of those works, but $\hat y$ in ours) clearly show an exponential decay in a broad region. In a very recent work~\cite{Amorosso:2026mdo} the characteristic length of such a decay was estimated to be $0.223(15) / \sqrt{\sigma}$, which is almost within the error from the value of $\lambda \sqrt{\sigma}$ quoted above.

\begin{table}[ht]
    \centering
    \begin{tabular}{|c|c|c|c|}
    \hline
    $T / T_c$ & $\beta$ & $a \, \sqrt{\sigma}$ & $\lambda \, \sqrt{\sigma}$ \\ \hline
    0.11      & 10.865  & 0.13137(86) & 0.25(4) \\ \hline
    \multirow{4}{*}{0.23} 
              & 8.768   & 0.16702(28) & 0.269(29) \\ 
              & 10.865  & 0.13137(86) & 0.245(35)  \\
              & 12.963  & 0.10870(62) & 0.227(58)  \\ 
              & avrg.   & -           & 0.256(29)  \\ \hline
    0.34      & 10.865  & 0.13137(86) & 0.235(18)  \\ \hline
    \end{tabular}
    \caption{Values of the intrinsic width at low temperatures obtained from fitting the exponential tails, in units of the string tension. For the intermediate value of the temperature, we averaged the three results obtained for different values of the lattice spacing.}
    \label{tab:ScaledLambdaExp}
\end{table}

The value can be compared with the inverse mass of the lowest lying glueball, $M_0 = 4.718(43) \, \sqrt{\sigma}$ (the value is quoted according to Ref.~\cite{Teper:1998te}), leading to the dimensionless product $\lambda \, M_0 = 1.21(11)$. This means that, at this level of precision, $\lambda$ is two standard deviations away from $1/M_0$. In the following parts of this section we will test how this estimate changes using more refined descriptions of the flux tube.

\subsection{Fitting with a convolution}
\label{sec:fit_convolutions}

Inspired by the results quoted in Section~\ref{sec:profile_u1} we fit the data assuming the same confining mechanism of the 3-dimensional $\mathrm{U}(1)$ model. Namely, we use the convolution of a Gaussian with a $\sech$ function:
\begin{equation}
\label{eq:conv_sech}
	\rho(y) = A^{\text{(conv)}}_1 \int_{-\infty}^{+\infty} \diff t \, 
	\exp \left( -\frac{t^2}{2 \, s^2} \right) \times
	\sech \left(\frac{y - t}{\lambda} \right) .
\end{equation}
The fit has three fitting parameters: $\lambda$, $s$ and the amplitude $A^{\text{(conv)}}_1$. Details on the implementation of this fit can be found in Appendix~\ref{app:sech} and the full results are reported in Tab.~\ref{tab:conv_fit_details}. 

We also modelled the data with an even simpler ansatz, assuming the profile of the flux tube as generated by an object with an intrinsic profile of the kind whose centre oscillates with Gaussian fluctuations. The resulting profile would be a convolution between the two, i.e.,
\begin{equation}
\label{eq:conv}
    \rho(y) = A^{\text{(conv)}}_2 \int_{-\infty}^{+\infty} \diff t \exp \left( -\frac{t^2}{2 \, s^2} \right) \times \exp(-|y - t| / \lambda).
\end{equation}
Details on the implementation of this fit can be found in Appendix~\ref{conv_exp}, corresponding results are collected in Tab.~\ref{tab:conv_fit_details2}.

Looking at these tables we see that the two convolutions give values of $\lambda$ which are in general compatible with each other; generally they also agree within the errors with the results of the simpler model discussed in Section~\ref{sec:fit_exp}. As expected the errors are smaller, but there seems to be no significant improvement in the values of the $\chi^2$.  The same considerations on the behaviour of $\lambda$ that we made at the end of the previous subsection hold also for these results.

From a qualitative point of view, the reasonable consistency of our data to these convolution models supports the picture of the flux tube as an object of non-vanishing width $\lambda$ that fluctuates generating a broader total width. However, given the limited statistical significance, we consider the output of these convolutional models as a term of comparison for other approaches, but not as reliable quantitative results. Let us, however, point out that even though we do not extract a consistent value for $\lambda$ at each lattice spacing, the optimal parameters of our fits are systematically larger than $1 / (a\, M_0)$, which is also reported in the caption of Tab.~\ref{tab:conv_fit_details}.

\subsection{The Clem Model}
\label{sec:fit_clem}

\begin{figure}[ht!]
    \centering
    \includegraphics[width=0.58333\linewidth]{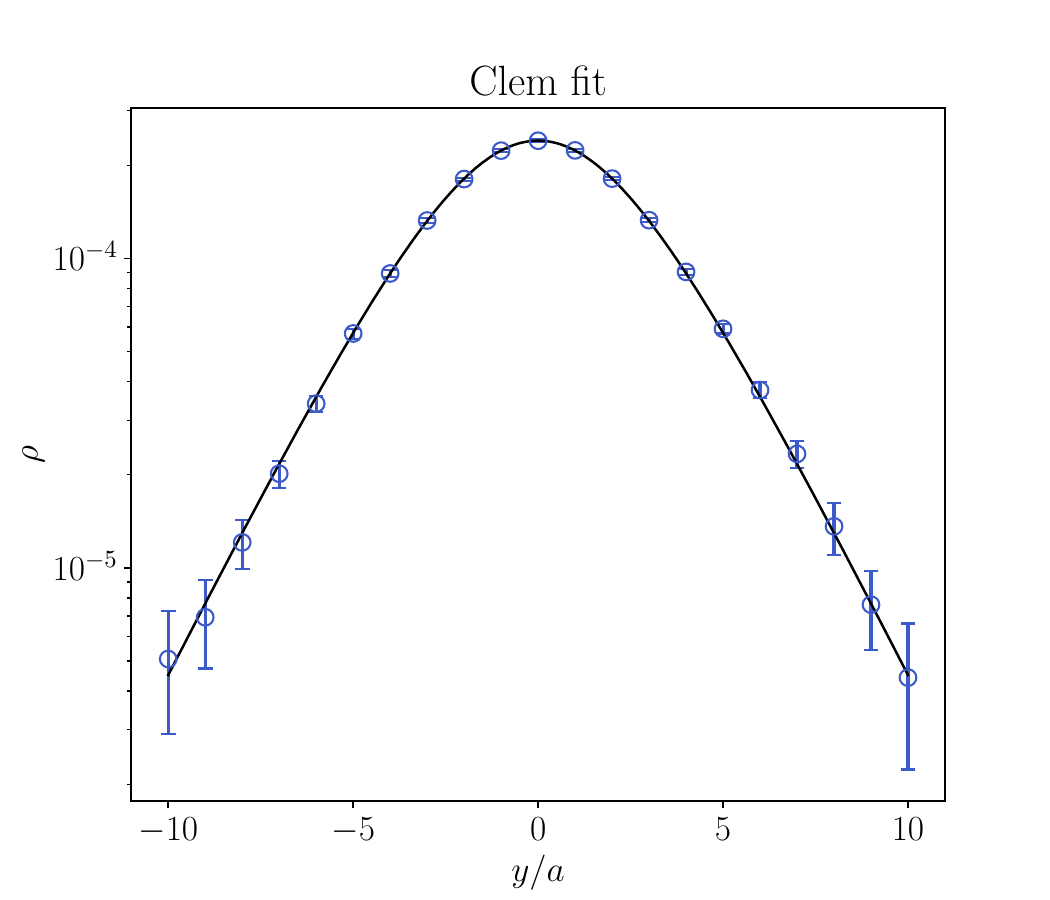}
    \caption{Example of a fit with the Clem model, see Eq.~\eqref{eq:clem}, for a profile obtained at $\beta = 10.865$ and $T = 0.23 \, T_c$, with $R = 11 a$.}
    \label{fig:clem_fit_example}
\end{figure}

Following the idea that the intrinsic width is a tool to explore the nature of the confining mechanism in the $\SU(2)$ gauge theory, we try to describe our data with the Clem model discussed in Section~\ref{sec:clem}, in particular fitting our numerical results for $\rho(y)$ for all available values of $y$ with Eq.~\eqref{eq:clem}, using $A^{\text{(Clem)}}$, $\xi$ and $\lambda$ as free parameters. 
The detailed results of all the fits can be found in Tab.~\ref{tab:clem_fit_details}, while a selection of results for $\xi$, $\lambda$, $w^2$ and $b$ are reported in Tab.~\ref{tab:clem_fit_example} and shown in Fig.~\ref{fig:clem_fit_example}. 

\begin{table}[ht!]
    \centering
    \textbf{Clem: fit results at $\beta = 10.865$} \\ [0.5em]
    \begin{tabular}{|c|c|c|c|c|c|}
    \hline
    $N_t$ & $R / a$ & $\lambda / a$ & $\xi / a$ & $w^2 / a^2$ & $b$ \\
    \hline
    60 &  9 & 1.95(15) & 3.13(22) & 9.98(67) & -0.538(62) \\
    \hline
    \multirow{4}{*}{30} 
       &  9 & 1.886(34) & 3.134(48) & 9.47(16) & -0.517(14) \\
       & 11 & 1.857(62) & 4.32(14) & 11.47(25) & -0.391(20) \\
       & 13 & 1.707(95) & 5.70(30) & 12.62(39) & -0.284(27) \\
       & 15 & 1.81(17)  & 6.42(59) & 14.83(66) & -0.271(43) \\
    \hline
    \multirow{4}{*}{20}
       &  9 & 1.801(37) & 3.335(60) & 9.24(15) & -0.473(14) \\
       & 11 & 1.768(83) & 4.55(19) & 11.15(36) & -0.359(28) \\
       & 15 & 1.83(26)  & 6.39(81) & 14.9(1.2) & -0.278(71) \\
       & 19 & 1.96(34)  & 7.7(1.3) & 18.6(1.7) & -0.255(83)\\
    \hline
    
    \end{tabular}
    \caption{Results from the fit assuming the Clem model to the profiles we collected at $\beta = 10.865$.}
    \label{tab:clem_fit_example}
\end{table}

The two parameters $\xi$ and $\lambda$ are particularly important since they allow to connect the dual superconductor picture with the flux tube one. As we mentioned in Section~\ref{sec:clem}, in the dual superconductor model they correspond to the two typical lengths of a vortex: $\xi$ is the radius of the core, while $\lambda$ is the London penetration length; their ratio determines the nature (type I or type II) of the superconductor. In the confining flux tube picture, $\lambda$ drives the exponential decay of the tails (and for this reason we identify it as the intrinsic width), while the product $\xi \lambda$ is the square width of the Gaussian peak.

Let us briefly comment on the results reported in Tab.~\ref{tab:clem_fit_details}. In general we find better $\chi^2$ values than those obtained with the convolutional models discussed in the previous subsection (with the same number of free parameters).  Given the good quality of these fits, we consider the Clem formula as our best parametrization of the profile of the flux tube at low temperature and the results for $\lambda$ reported in Tab.~\ref{tab:clem_fit_details} as our best estimates for the intrinsic width.

\begin{figure}[ht]
    \centering
    \includegraphics[width=\textwidth]{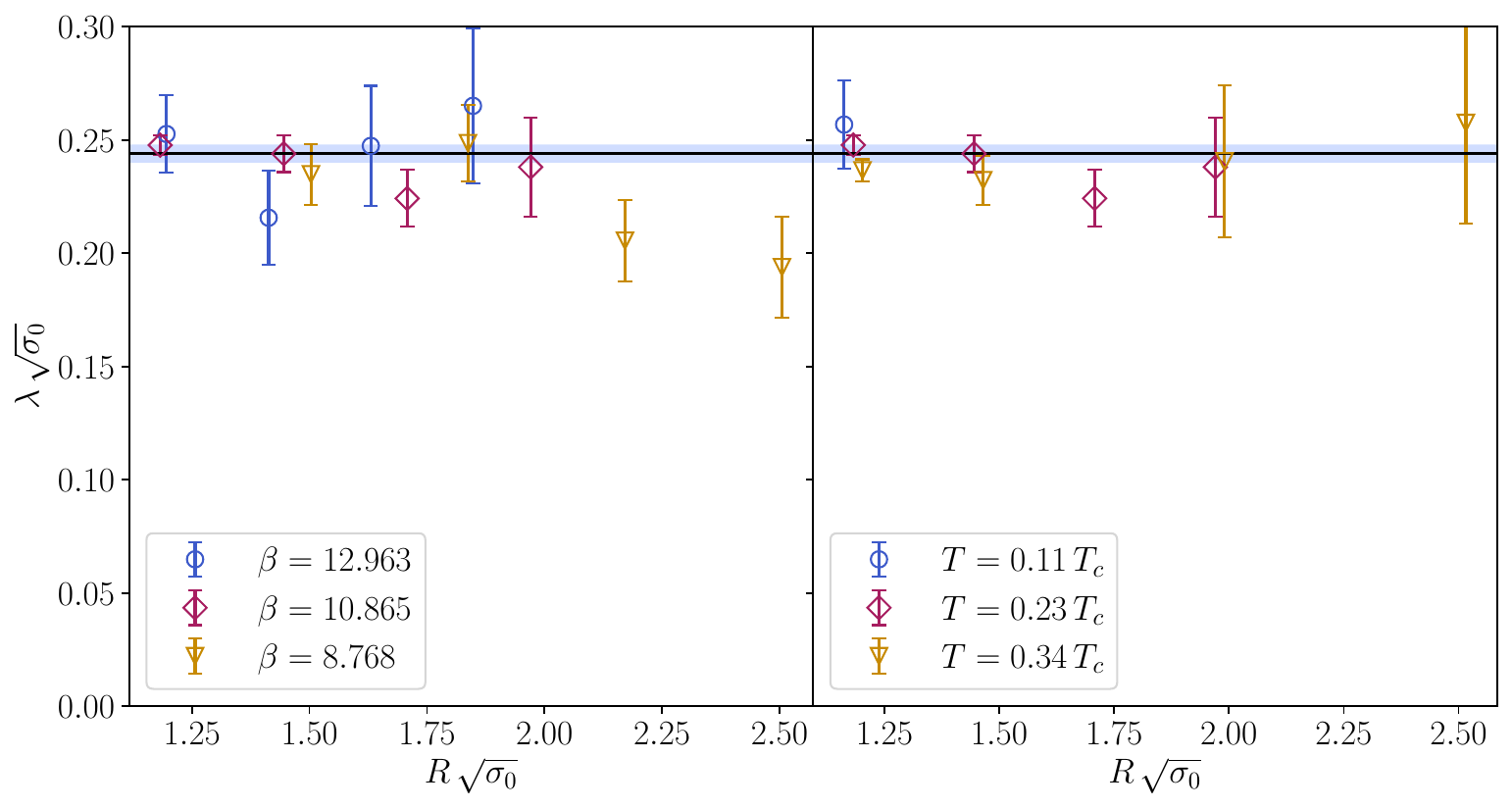}
    \caption{Different values of $\lambda$ in units of the string tension, showing no dependence on the distance between the sources $R$. In the left panel the temperature is fixed to $T = 0.23 \, T_c$ and we compare different values of the lattice spacing. In the right panel we show different temperatures keeping the lattice spacing fixed by setting $\beta = 10.865$. The solid line and the shaded region correspond to our final estimation and its uncertainty.}
    \label{fig:ScaledLambdaLowT}
\end{figure}

Determinations of $\lambda$ with the Clem model appear to be stable when changing $R$. In order to check for a possible dependence of $\lambda$ on the lattice spacing, we focused on the case of $T = 0.23 \, T_c$, where we have three different values of the lattice spacing. For each $a$ we took a weighted average of $\lambda$ at different distances, that we report in Tab.~\ref{tab:ScaledLambdaLowT} for $\lambda \sqrt{\sigma}$. The values at the three different spacings are compatible within the errors: we take their average as our final estimation of the intrinsic width at that temperature, which turns out to be\footnote{Even if this result has a much smaller error than the one quoted in Section~\ref{sec:fit_exp}, it still displays only a mild tension (just above one standard deviation) with the value obtained in Ref.~\cite{Amorosso:2026mdo}.} $\lambda \sqrt{\sigma}=0.244(4)$.
Looking at Tab.~\ref{tab:ScaledLambdaLowT} we see that the values obtained at the other two temperatures (for which we have only one lattice spacing) are also compatible with this result, suggesting no dependence on $T$ in the low temperature region. Looking at figure \ref{fig:ScaledLambdaLowT}, we can appreciate the stability changing $R, T$ and $\beta$ (or equivalently the lattice spacing). 

Our final value for $\lambda\sqrt{\sigma}$ is at the same scale of the inverse of the lowest glueball mass of the gauge theory, but largely incompatible with its value: for the $\SU(2)$ pure gauge theory in $(2+1)$ dimensions, Ref.~\cite{Teper:1998te} provides $M_0 / \sqrt{\sigma} = 4.718(43)$, leading to $\lambda \, M_0 = 1.16(3)$.  It would be interesting to understand if this disagreement is due to the non-abelian nature of the flux tube or is instead a signature of the lack of consistency of the dual superconductor model of confinement.

Indeed, our study leads us to believe that the Clem model is a good starting point, but most likely not the final answer: in particular,  the $\xi$ scale, according to our fits, is clearly a growing function of $R$. On one hand, this behaviour is perfectly consistent with the EST prediction of a logarithmic broadening of the flux tube; on the other hand, it leads to a peculiar scenario in the dual superconductor picture. If we use the formula by Clem, reported in Eq.~\eqref{eq:clem_ginsland}, to compute the Ginzburg--Landau parameter, we find that it decreases with the distance $R$ between the Polyakov loops. 
If we are to assume the logarithmic broadening to hold at large distances and the intrinsic width to stay constant, we conclude that the length $\xi$ needs to approach infinity at large $R$ and consequently $\kappa$ would tend to zero.

Although all our results lay in the region corresponding to a type I dual superconductor ($\kappa < 1 / \sqrt{2}$), we cannot exclude that this does not apply at shorter distances. Interestingly, the critical value of the Ginzburg--Landau value could be reached for $R \simeq 2 R_c$, \footnote{$R_c$ is the scale below which the EST model is not any more valid, see below for a precise definition.} extrapolating our data under a logarithmic broadening assumption. 
It is possible that this surprising behaviour could play a role in the uncertainty regarding the type of dual superconductor determined in previous works on $\SU(3)$ in $(3+1)$ dimensions~\cite{Cea:2012qw,Cardoso:2010kw}.

\begin{table}[ht]
    \centering
    \begin{tabular}{|c|c|c|c|}
    \hline
    $T / T_c$ & $\beta$ & $a \, \sqrt{\sigma}$ & $\lambda \, \sqrt{\sigma}$ \\ \hline
    0.11      & 10.865  & 0.13137(86) & 0.26(2) \\ \hline
    \multirow{4}{*}{0.23} 
              & 8.768   & 0.16702(28) & 0.240(13) \\ 
              & 10.865  & 0.13137(86) & 0.245(5)  \\
              & 12.963  & 0.10870(62) & 0.242(17) \\ 
              & avrg.   & -           & 0.244(4)  \\ \hline
    0.34      & 10.865  & 0.13137(86) & 0.236(5)  \\ \hline
    \end{tabular}
    \caption{Values of the intrinsic width at low temperatures obtained from fits using the Clem formula, see Eq.~\eqref{eq:clem}, in units of the string tension. For the intermediate value of the temperature, we averaged the three results obtained for different values of the lattice spacing.}
    \label{tab:ScaledLambdaLowT}
\end{table}

\subsubsection{On the logarithmic growth of the total flux tube width}

\begin{figure}
    \centering
    \includegraphics[width=0.58333\textwidth]{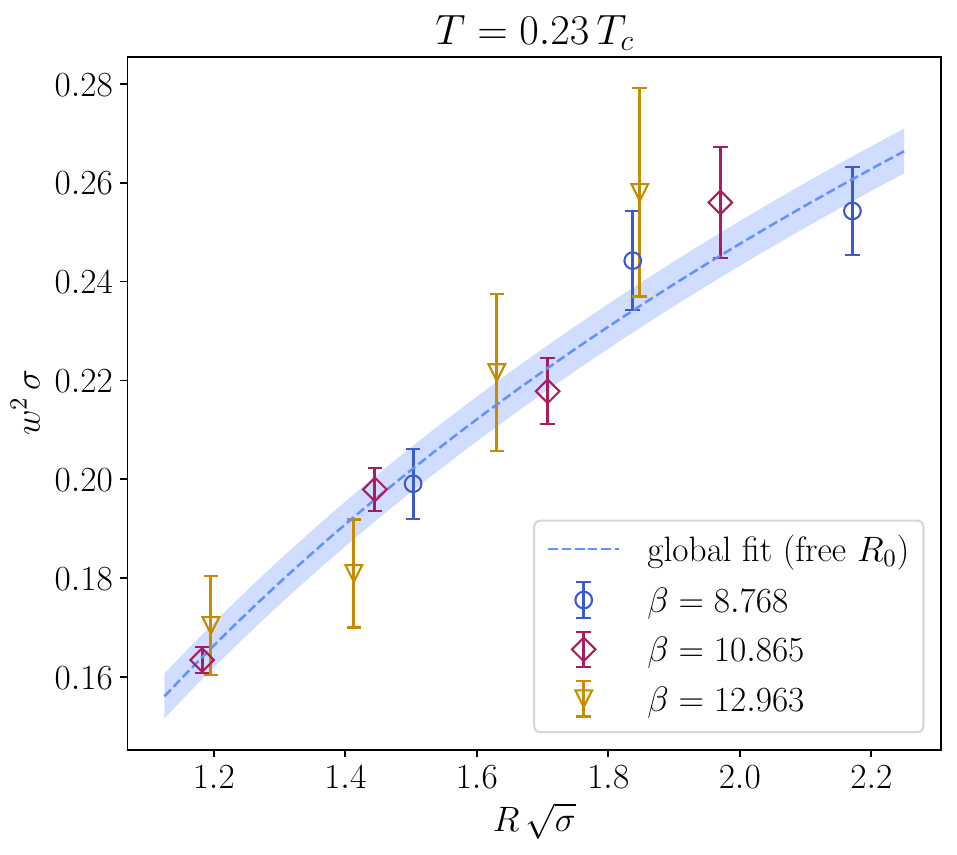}
    \caption{The logarithmic broadening of the flux tube increasing its length $R$. The data-point are the values we found at $T = 0.23 \, T_c$. The dashed line is a fit to all the plotted data according to Eq.~\eqref{eq:EST_logbroad} with $R_0$ as the only free parameter. The coefficient multiplying the logarithm was fixed interpolating the values of the string tension in Ref.~\cite{Athenodorou:2016ebg}.}
    \label{fig:log_broad}
\end{figure}

Assuming the Clem ansatz is a valid quantitative description for the shape of the flux tube it is possible, from the fitted values of $\lambda$ and $\xi$, to compute numerically the total width $w^2$ of the flux tube and test the corresponding EST predictions. This procedure allows to obtain excellent estimates of $w^2$, much more precise than those obtained with a naive numerical integration of the profile (see Section~\ref{sec:fit_exp}), leading to an accurate test of the logarithmic broadening of $w^2$ predicted by EST.

We fitted our values for $w^2$ using Eq.~\eqref{eq:EST_logbroad} and fixing $1 / (2 \pi \sigma)$ to the values obtained interpolating the results for $\sigma$ in Ref.~\cite{Athenodorou:2016ebg}. This leaves the radius $R_0$ as the only free parameter of the model. The data turn out to be in remarkable agreement with this description: the fit shown in Fig.~\ref{fig:log_broad}, including all our data point at $T = 0.23 \, T_c$ at different lattice spacings, has a reduced $\chi^2$ of $0.56$ with a final result of $R_0 \sqrt{\sigma} = 0.422(12)$. 

Interestingly, this determination of $R_0$ is in disagreement with the value $R_0 \sqrt{\sigma} = 0.364(3)$ that was found in Ref.~\cite{Gliozzi:2010zv} assuming a different phenomenological profile for the flux tube that does not take into account the intrinsic width.

\subsection{The flux tube profile at high temperature}
\label{sec:fit_sy}

We now focus on results obtained at high temperature, but still for $T<T_c$; the simulation setup is reported in Tab.~\ref{tab:DetailsHighT}.
In this region, the quality of the fits assuming the convolution models or the Clem formula becomes much worse and the value of $\lambda$ that we extract from those fits is not uniform any more for different values of $R$. This behaviour clearly suggests that we are beyond the regime where such approaches are applicable, and a different description is needed: thus, we turn to the Svetitsky--Yaffe mapping.

\subsubsection{The Svetitsky--Yaffe mapping for the flux tube profile}

We report the fit results of our high temperature data with Eq.~\eqref{eq:SY_profile} in Tab.~\ref{tab:ScaledLambdaHighT}, where we can appreciate the good quality of the fits.
The fact that the $A^{\rm (SY)}$ and $\lambda$ parameters are approximately constant for different values of the distance $R$ on the same lattice further corroborates this approach.

Given this consistency, we attempted a global fit of all the profile data from the each lattice at different values of $R$, whose full results can also be found in Tab.~\ref{tab:symp_fit_details} and~\ref{tab:symp_fit_details2}. These fits show the high predictive power of the SY mapping, which uses only two free parameters for $\mathcal{O}(100)$ data points. Considering only large enough values of $R$, we still obtain satisfying values for the $\chi^2$ and stable values of the fitting parameters. We report our final values for several lattice spacings at two different temperatures in Tab.~\ref{tab:ScaledLambdaHighT}, where we have converted the results in units of the string tension using again Ref.~\cite{Athenodorou:2016ebg} for the scale setting.

\begin{table}[htb]
    \centering
    \begin{tabular}{|c|c|c|c|c|c|}
    \hline
    $T / T_c$ & $\beta$ & $a \, \sqrt{\sigma}$ & $\lambda \, \sqrt{\sigma}$ & $\sqrt{\sigma} / (2E_0)$ \\
    \hline
    \multirow{4}{*}{0.68}
         & 10.865 & 0.13137(86) & 0.608(18) & \multirow{4}{*}{0.597} \\ 
         & 11.914 & 0.11881(81) & 0.677(31) & \\
         & 12.962 & 0.10486(51) & 0.578(21) & \\
         & 14.011 & 0.10046(35) & 0.560(25) & \\ \hline
    \multirow{2}{*}{0.85}
         & 10.865 & 0.13137(86) & 1.332(32) & \multirow{2}{*}{1.298} \\
         & 13.424 & 0.10870(62) & 1.519(80) & \\ \hline
    \end{tabular}
    \caption{Values of the intrinsic width at high temperature and comparison with the ground state of the string computed as in Ref.~\cite{Caristo:2021tbk}. At $T = 0.49 \, T_c$ the SY based model does not fit well the data, hence our estimation relayed on the method explained in Section \ref{sec:fit_exp}. Note that at that temperature, the ground state of the closed string is an extrapolation, and to be directly measured would need very long strings.}
    \label{tab:ScaledLambdaHighT}
\end{table}

We can further test the SY mapping by checking the compatibility between the value of $\lambda$ obtained fitting with Eq.~\eqref{eq:SY_profile} and the value derived from the ground state energy of the effective string. For this reason in Tab.~\ref{tab:ScaledLambdaHighT} (and more extensively in in Tab.~\ref{tab:symp_fit_details} and~\ref{tab:symp_fit_details2}) we also report the value of $1 / (2 E_0)$, which in turn was estimated from the exponential decay of the two-point Polyakov loop correlator at large $R$, namely $\exp(-E_0 \, R)$. The value of $E_0$ was estimated from a fit analogous to those described in Refs.~\cite{Caselle:2012rp, Caselle:2024zoh}, to data obtained on the same configurations as those used to compute the profile. We find that the estimation of $\lambda$ from the profile is always compatible within three standard deviations with the value evaluated from the $\braket{P P^\dagger}$ correlator.

\begin{figure}[ht]
    \centering
    \includegraphics[width=0.583333\textwidth]{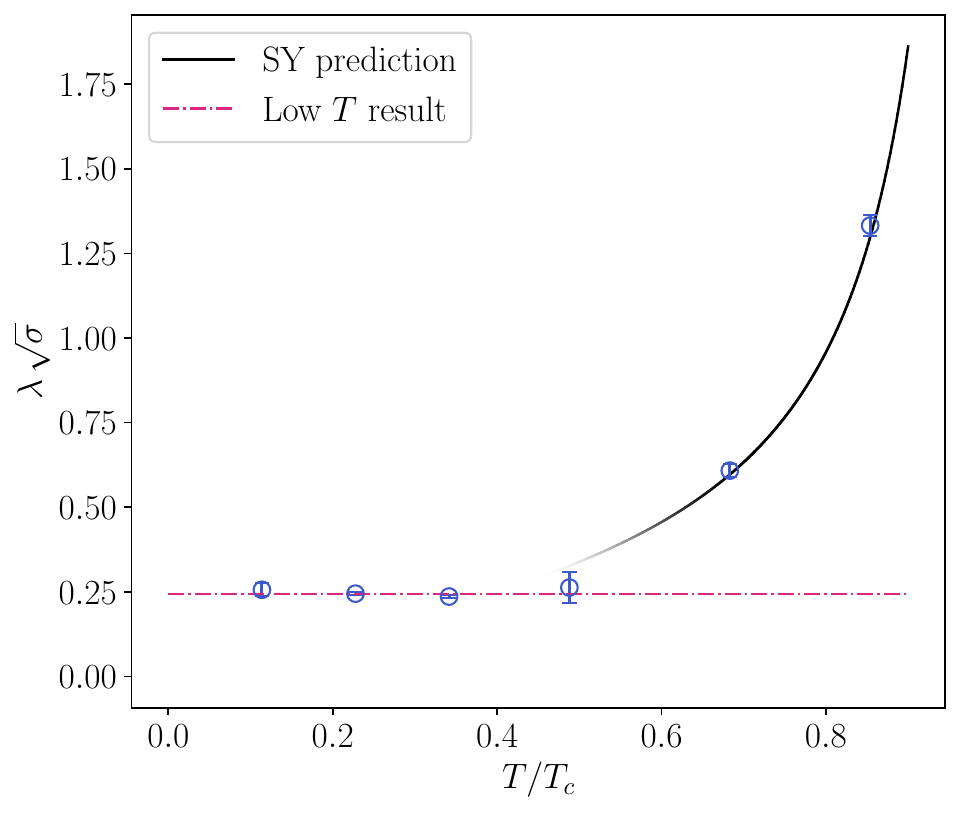}
    \caption{The values of $\lambda$ as a function of the temperature. The red dash-dotted line is our estimation at low temperature. The black solid line is the SY prediction, using the fit from Ref.~\cite{Caselle:2024zoh} for $E_0$.}
    \label{fig:lamb_vs_temp}
\end{figure}

We remark that these tests of the SY mapping are successful even at temperatures as small as $T = 0.68 \, T_c$, allowing us to use to predict the behaviour of the flux tube profile in a large range of temperatures. In Fig.~\ref{fig:lamb_vs_temp} we plot the dependence of $\lambda$ on the temperature, immediately showing its constant value at low $T$. 
Assuming $\lambda = 1 / (2 E_0)$ and using the fit from Ref.~\cite{Caselle:2024zoh} to compute $E_0$ we are also able to describe remarkably well our data points with a curve representing the SY prediction at high $T$; since $E_0$ is supposed to vanish at the deconfinement transition, the intrinsic width consistently diverges approaching $T_c$. Note that, since at $T = 0.49 \, T_c$ neither the Clem nor the SY based model give satisfactory results, we use the value from the exponential fit of the tails, as described in Section~\ref{sec:fit_exp}.

\begin{figure}
    \centering
    \includegraphics[width=\textwidth]{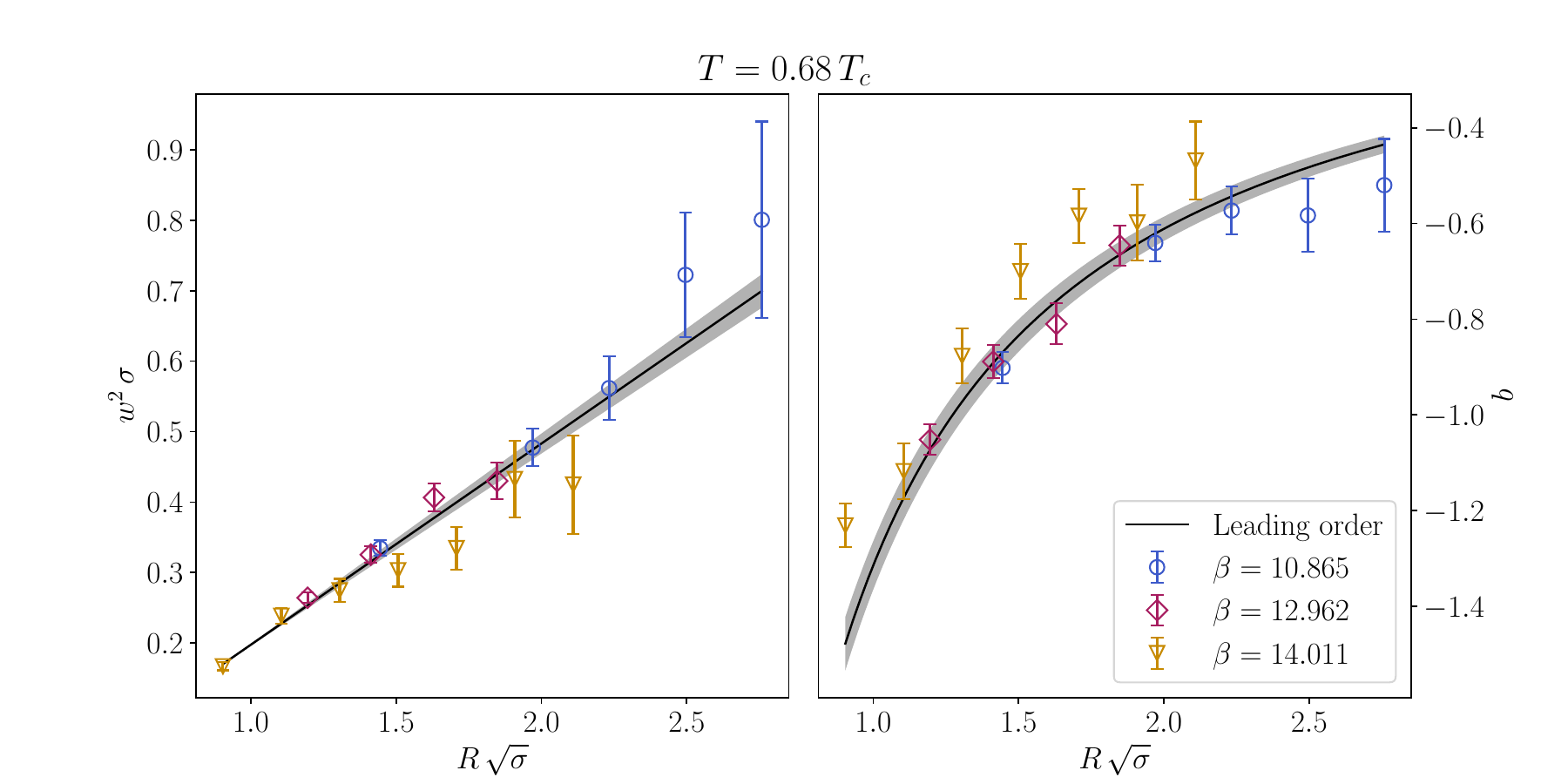}
    \caption{Square of the total width and Binder cumulant from the fit assuming the SY mapping of our data at $T = 0.68 \, T_c$. The solid line with the confident band is the expansion from Eq.~\eqref{eq:SY_linearbroad} in the left panel and Eq.~\eqref{eq:SY_binder} in the right one. The input value of $\lambda$ is the one extracted from the Polyakov loop correlator $\lambda = 1 / (2 E_0)$.}
    \label{fig:compare_binderwidt_07tc}
\end{figure}

Furthermore, we can also see how the SY mapping agrees with the prediction of a linear broadening of the flux tube at high temperature. It is, indeed, possible (see Appendix~\ref{app:moments_SY} for more details) to analytically compute the moments of the SY profile (as a function of $\lambda$ and $R$) and in particular to expand the total width (squared) $w^2$ of the flux tube:
\begin{equation}
    \label{eq:SY_linearbroad}
    w^2(R) = \frac{\lambda R}{2} - \frac{\lambda^2}{2} + \mathcal{O}\left( \frac{\lambda^3}{R} \right) .
\end{equation}
In this expansion the leading order is linear in $R$: although this rigorously hold in the limit $R \gg 2 \lambda$ (which is numerically difficult to probe), since the natural expansion parameter is $2 \lambda / R$, it is already appreciable from our data, as shown in the left panel of Fig.~\ref{fig:compare_binderwidt_07tc}.

Then, we can also compare the coefficient of this term to the one in Eq.~\eqref{eq:EST_linbroad}. We have already numerically checked the relation $\lambda = 1 / (2 E_0)$: in the EST language at high-$T$, it is natural to relate $E_0$ to the (temperature-dependent) string tension by $E_0 = \sigma(T) / T$.
Recalling $L_t = 1 / T$, we can rewrite Eq.~\eqref{eq:SY_linearbroad} as
\begin{equation}
    w^2(R) = \frac{R}{4 \, L_t \, \sigma(T)} + \dots \, ,
\end{equation}
which perfectly matches the leading term in Eq.~\eqref{eq:EST_linbroad}. Notice, as a side remark, that with this mapping we exactly find at the denominator the temperature dependent string tension of Eq.~\eqref{eq:conjecture}  thus giving further support to the conjecture of Ref.~\cite{Caselle:2010zs}.


Finally, we can also compute the Binder cumulant of the flux tube profile, assuming the SY mapping, and notice that its leading term is negative and proportional to $1/R$:
\begin{equation}
    b = -\frac{2 \, \lambda}{R} + \mathcal{O} \left( \frac{\lambda^2}{R^2} \right).
    \label{eq:SY_binder}
\end{equation}
We show the estimation of the Binder cumulant from the fits assuming the SY mapping in the right panel of Fig.~\ref{fig:compare_binderwidt_07tc}.

\section{Conclusions}
\label{sec:conclusions}

In this work we performed a numerical study of the profile of the flux tube between two static sources in $\SU(2)$ Yang-Mills theory in $(2+1)$ dimensions. Through a careful analysis of the results, a typical length scale of the theory clearly emerged, which we identified as the intrinsic width $\lambda$ of the flux tube. Such a scale is independent on the length of the flux tube and stable at low temperatures. The main manifestation of this intrinsic width lies in the exponentially decaying tails in the flux tube profile. These tails constitute an evident deviation from the Gaussian profile, which in turn would be typical of widthless oscillating objects. We estimated the value of the intrinsic width in units of the string tension to be $\lambda \sqrt{\sigma}= 0.244(4)$, which interestingly is essentially compatible with the value extracted in Ref.~\cite{Amorosso:2026mdo} from the entanglement entropy of the flux tube. 

The relationship with the inverse of the mass of the lightest glueball is of particular relevance: we find that $\lambda$ is indeed in the same ballpark of $1/ M_0$, but statistically incompatible with its precise value. This issue deserves a thorough investigation, since both in the $\mathrm{U}(1)$ case and in the holographic analysis of Refs.~\cite{Danielsson:1998wt, Canneti:2025afi} the intrinsic width equals exactly the inverse of the mass gap of the theory. Our future goals include a careful numerical determination of $\lambda \sqrt{\sigma}$ in $\SU(N)$ gauge theories with increasing $N$, in order to understand how the results of careful lattice computations approach the holographic large-$N$ prediction.

Besides estimating the intrinsic width and arguing that it is an intrinsic property of the gauge theory (in the same way deviations from the Nambu-Got\=o Effective String Theory are), we provided further quantitative statements regarding the profile of the flux tube. We focused especially on its non-Gaussianity, which we parametrized using a combination of moments equivalent to the Binder cumulant: we found that the latter is negative and decreasing in modulus when the flux tube increases in length. To do so we examined different possible parametrizations for the profile of the flux tube: at low temperatures we found that the most reliable description is given by the Clem formula, which arises in the context of the dual superconductive model for confinement. Interestingly, in our analysis taking the dual superconductor picture at face value leads to a possible inconsistency: in particular, we observe that the Ginzburg--Landau parameter we extract from our analysis of the profile shows a clear dependence on flux tube length. It would be interesting to test whether refined versions of confining models based on monopole condensation (for example those recently discussed in Refs.~\cite{Nguyen:2024ikq,Nguyen:2025voy,Domurcukgul:2025xgf,Giansiracusa:2025hfj}) exhibit the same behaviour. More generally, we believe that our high precision numerical results may serve as a useful benchmark for testing new or existing models of confinement.

We extended our analysis of the non-Gaussianity of the flux tube profile in the high temperature regime, approaching the deconfinement phase transition from below. In this regime we once more found a scale that parametrizes the exponentially decaying tails of the profile, which does not depend on the length of the flux tube. In this case, it has a clear dependence on the temperature that can be fully understood exploiting the Svetitsky--Yaffe mapping: namely, the latter predicts in the proximity of the deconfinement transition that the intrinsic width $\lambda$ is proportional to the inverse of the ground state energy of the effective string. This is perfectly consistent with an intrinsic width that diverges approaching the critical temperature.

The Svetitsky--Yaffe mapping leads to a model for the flux tube profile which is in very good agreement with our data, even at a temperature as small as $T = 0.68 \, T_c$. The model is indeed impressively predictive, allowing to fit all the profiles obtained on a lattice (given that the flux tube is long enough) with only two parameters, and working well on a wide range of temperatures. This work thus adds one further impressive test to a long list of studies~\cite{Caselle:2006wr, Caselle:2010zs, Caselle:2012rp, Caselle:2017xrw, Caristo:2021tbk, Caselle:2024zoh, Caselle:2025elf} of the Svetitsky--Yaffe mapping for the $\SU(N)$ Yang--Mills theories in $(2+1)$ and $(3+1)$ dimensions.

\vskip 1.5cm
\noindent {\large {\bf Acknowledgments}}
\vskip 0.2cm
We thank Rocco Amorosso, Claudio Bonati, Tommaso Canneti and Marco Panero for several insightful discussions. All authors acknowledge support from the SFT Scientific Initiative of INFN. The work of M.~C., A.~N., D.~P. and L.~V.~was supported by the Simons Foundation grant 994300 (Simons Collaboration on Confinement and QCD Strings). A.~N.~acknowledges support by the European Union - Next Generation EU, Mission 4 Component 1, CUPD53D23002970006, under the Italian PRIN “Progetti di Ricerca di Rilevante Interesse Nazionale – Bando 2022” prot. 2022ZTPK4E.
\vskip 1cm

\appendix

\section{Details on fitting the exponential tails}
\label{app:details_exp_fit}

\begin{figure}[ht!]
    \centering
    \includegraphics[width=0.58333\textwidth]{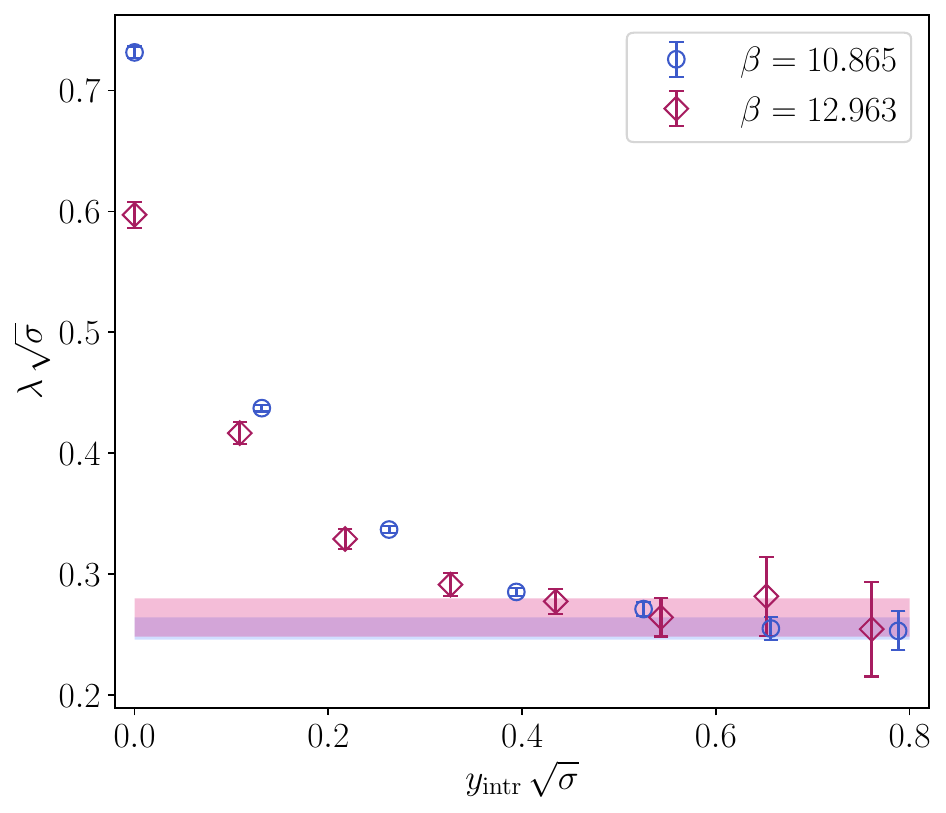}
    \caption{Values of $\lambda$ obtained fitting with an exponential the tails of the profile using different values of $y_\intr$. We compare  $R / a = 9$ on the $\beta = 10.865$, $N_t = 30$ lattice with $R / a = 11$ on the $\beta = 12.963$, $N_t = 36$. For small values of $y_\intr$, $\lambda$ is considerably overestimated. The shaded regions correspond to the values of the identified plateau.}
    \label{fig:yint_choice}
\end{figure}

In section \ref{sec:fit_exp} we separated in the profile a Gaussian peak region $-y_\intr < y < y_\intr$ from two exponential tails $|y| > y_\intr$ and fitted the profile in the latter. In this procedure, it is extremely important to assess the effect of the choice of $y_\intr$, which can be fully appreciated in Fig.~\ref{fig:yint_choice}. In particular, since the profile (at least near the peak) is logarithmically concave ($(\diff / \diff y)^2 \log \rho(y) < 0$), choosing $y_\intr$ to be too small leads to an overestimation of the parameter $\lambda$, with a consequent, uncontrolled, systematic overestimation of every integral. 
On the other hand, assuming that for large enough transverse displacement the exponential decay is the only dominant effect in the profile, choosing $y_\intr$ too large will not lead to systematic effects. Indeed, we expect the values of $\lambda$ obtained with larger and larger values of $y_\intr$ to be consistent and thus the values of the integrals to be compatible within the errors. However, due to the loss of relative precision in the data points at large values of $y$, choosing a very large $y_\intr$ leads to an increase of the statistical error on the parameter $\lambda$ and consequently on our final results, see Fig.~\ref{fig:yint_choice}. For this reason, we compared the results for different choices of $y_\intr$, and we selected the first value large enough to be in the stable region for $\lambda$.

In order to estimate the statistical error, we performed a bootstrap analysis, repeating the computations on $500$ resampled profiles, extracted from a multivariate Gaussian distribution, centred around the mean value of the numerical data and with the same covariance matrix. Variance and covariance of the data were numerically extracted from the Monte Carlo samples via a $\Gamma$-method analysis~\cite{Wolff:2003sm,Joswig:2022qfe}. Notice that all the pieces of the integrals (in particular the contributions from the tails and the central interval) are computed on the same resampled profiles. The standard deviation over the bootstrap sample is taken only after the final observables are computed, in order to take into account the correlation between different contributions to the integrals.

\subsection{Note on \texorpdfstring{$y\textsubscript{intr}$}{y intr}}
\label{determination_of_y}

\begin{figure}[htb]
    \centering
    \includegraphics[width=0.5833333\linewidth]{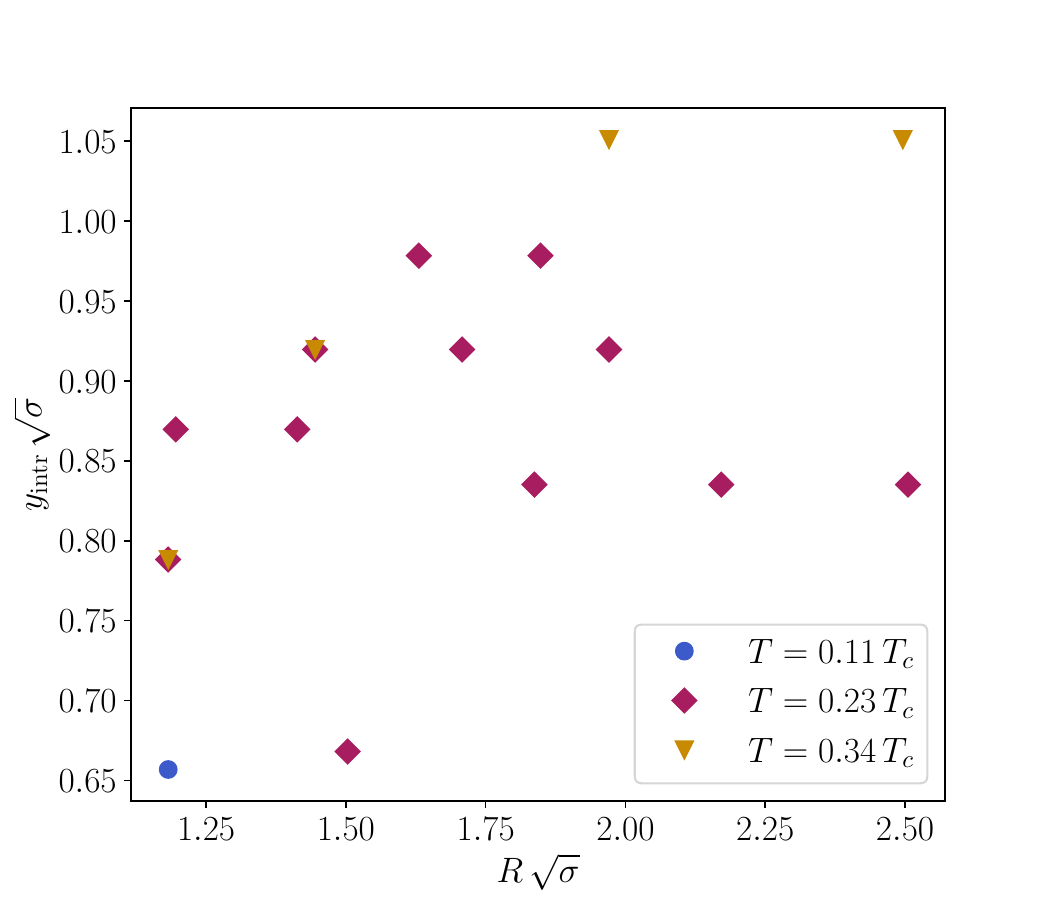}
    \caption{Optimal values of $y_\intr$ for the profile we measure, in units of the string tension, plotted against the distance between the Polyakov loops in the same units. Although a general increasing trend is evident, its interpretation as a physical length should be carefully considered.}
    \label{fig:compare_y_intr}
\end{figure}

It is not obvious from our data if the best value of $y_\intr$ can be predicted from the simulation parameters or not. It appears to be an increasing function of the distance between the Polyakov loops $R$ and to increase in the continuum limit (as a physical length should). The values of $y_\intr$ in units of the string tension plotted against the distance between the Polyakov loops are shown in Fig.~\ref{fig:compare_y_intr}. However, being deduced from a plateau in plots like the one shown in Fig.~\ref{fig:yint_choice}, it is possible that the precision of the data plays an important role in the final result.

\section{Details on the convolution implementation}

\subsection{Using the \texorpdfstring{$\sech$}{sech} function}
\label{app:sech}

Since this integral of Eq.~\eqref{eq:conv_sech} cannot be reduced to functions we can easily find tabulated, we computed it numerically, by truncating the $t$-integral in the interval $[-L, L]$, where we chose $L = 10 \, s$. The truncation is justified by the rapid decay of the Gaussian and we checked that the results of the fit would not be altered if we chose $L = 5 \, s$ instead. The numerical integration is performed with the function \textit{quad\_vec}, in the \textit{scipy.integrate} python package. Our fitting model is thus
\begin{equation}
\rho(y) = \tilde A^\text{(conv)} \, \int_{-L}^{+L} \diff t \; \exp \left( -\frac{t^2}{2 \, s^2} \right) \times \sech \left( \frac{y - t}{\lambda} \right).
\end{equation}

The  moments of the profile, instead, can be computed analytically. In particular, it suffices to recall that the moment-generating function of a convolution is the product of those of the convolved distributions (or, equivalently, the Laplace transform of a convolution is the product of the Laplace transforms of the factors). In our case it is, thus, easy to write down the Taylor series of the moment-generating function $z(t)$:
\begin{multline}
	z(t) =
	\exp\left(\frac{s^2 \, t^2}{2}\right) \,
	\sec \left( \frac{\pi \, t \, \lambda}{2} \right) = \\
	1 + \left(\frac{\pi ^2 \lambda ^2}{8}+\frac{s ^2}{2}\right) t^2
	+ \left(\frac{5 \pi ^4 \lambda ^4}{384}+
	\frac{1}{16} \pi ^2 \lambda ^2 s ^2+\frac{s ^4}{8}\right) t^4
	+ \mathcal{O}\left(t^6\right)
\end{multline}
From $z(t)$ we can extract the relevant moments in the following way:
\begin{align}
	w^2 &= \left. \left( \frac{\diff}{\diff t} \right)^2 \, z(t) \right|_{t = 0} =
	s^2 + \frac{\pi^2}{4} \lambda^2 \\
	\mu_4 &= \left. \left( \frac{\diff}{\diff t} \right)^4 \, z(t) \right|_{t = 0} =
	4! \, \left( \frac{5 \pi ^4 \lambda ^4}{384}+\frac{1}{16} \pi ^2 \lambda ^2 s ^2+\frac{s ^4}{8} \right).
\end{align}
Finally, we can also compute the Binder cumulant according to Eq.~\eqref{eq:binder_def} as:
\begin{equation}
	b = -\frac{2}{3} \left( \frac{1}{1 + \left( \frac{2 \, s}{\pi \, \lambda} \right)^2} \right)^2,
\end{equation}
which, we observe, is negative by construction. It goes to zero in the limit where the ratio $s / \lambda$ is large, where the model approaches a Gaussian.

\subsection{Using the \texorpdfstring{$\exp$}{exp} function}
\label{conv_exp}

\begin{figure}[ht]
    \centering
    \includegraphics[width=0.58333\linewidth]{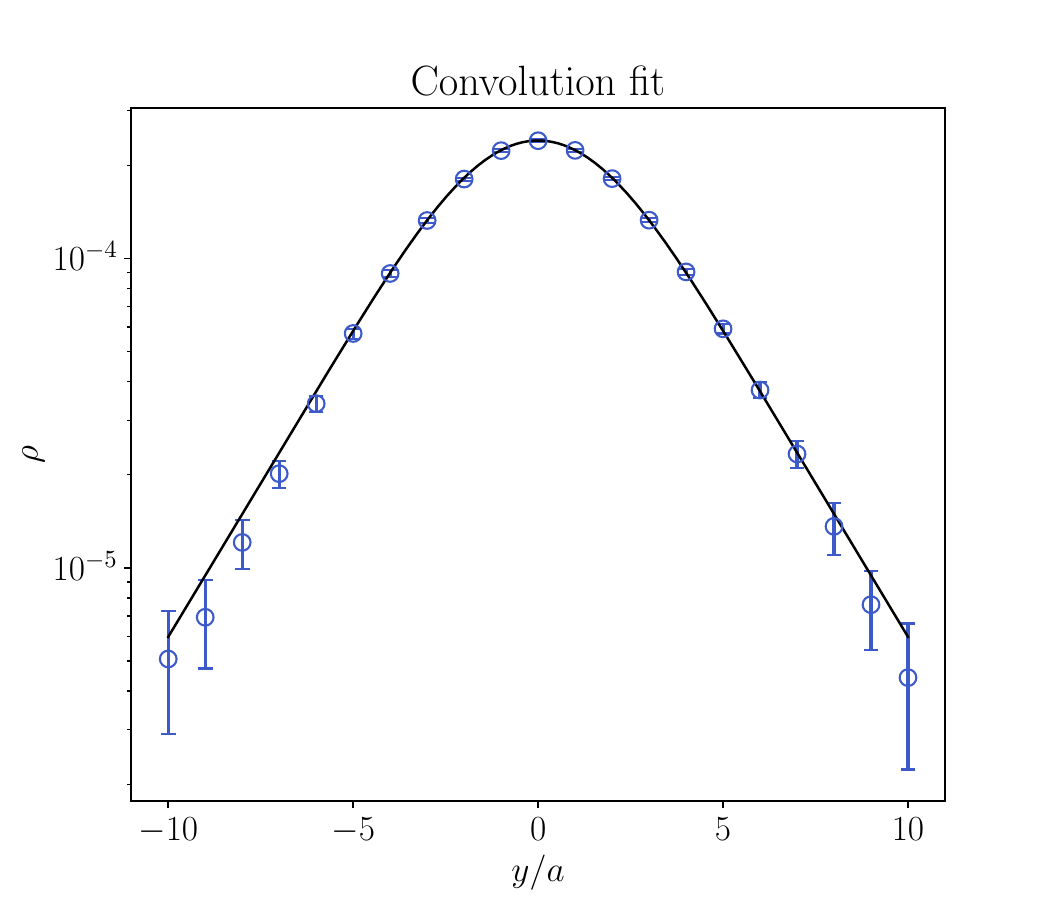}
    \caption{Example of a fit with the convolution model, see Eq.~\eqref{eq:conv}, for a profile obtained at $\beta = 10.865$ and $T = 0.23 \, T_c$, with $R = 11 a$.}
    \label{fig:conv_fit_example}
\end{figure}

Differently from Eq.~\eqref{eq:conv_sech}, the integral in Eq.~\eqref{eq:conv} can be expressed in terms of the complementary error function $\erfc(x)$, defined as
\begin{equation}
    \erfc(x) = \frac{2}{\sqrt{\pi}} \int_x^\infty \diff t \, \exp(-t^2),
\end{equation}
finding the following explicit result:
\begin{equation}
    \rho(y) = A^\text{(conv)} \left( e^{-y / \lambda} \, \erfc\left( \frac{s}{\sqrt{2} \, \lambda} - \frac{y}{\sqrt{2} \, s}\right) + e^{y / \lambda} \, \erfc\left( \frac{s}{\sqrt{2} \, \lambda} + \frac{y}{\sqrt{2} \, s}\right)\right) .
\end{equation}
An example of a fit done with this function is shown in Fig.~\ref{fig:conv_fit_example}.

The most convenient way of computing the moments is to proceed as in the previous case, multiplying the moment-generating functions of the two convolved factors:
\begin{multline}
    z(t) =
    \exp\left(\frac{s^2 \, t^2}{2}\right) \, \frac{1}{1 - \lambda^2 \, t^2} = 
    \left(1 + \frac{s^2 \, t^2}{2} + \frac{s^4 \, t^4}{8} + \dots \right) 
    \left(1 + \lambda^2 \, t^2 + \lambda^4 \, t^4 + \dots\right) = \\
    = 1 + \left( \frac{s^2}{2} + \lambda^2 \right) \, t^2 + 
    \left( \frac{s^4}{8} + \frac{s^2 \, \lambda^2}{2} + \lambda^4 \right) \, t^4 + \dots \; .
\end{multline}
From $z(t)$ we read the moments:
\begin{align}
    w^2 &= \left. \left( \frac{\diff}{\diff t} \right)^2 \, z(t) \right|_{t = 0} =
    s^2 + 2 \, \lambda^2 \\
    \mu_4 &= \left. \left( \frac{\diff}{\diff t} \right)^4 \, z(t) \right|_{t = 0} =
    4! \, \left( \frac{s^4}{8} + \frac{s^2 \, \lambda^2}{2} + \lambda^4 \right)
\end{align}
and finally compute the Binder cumulant:
\begin{equation}
    b = - \left( \frac{1}{1 + \tfrac{1}{2} (s / \lambda)^2} \right)^2.
    \label{eq:binder_conv}
\end{equation}

\section{Moments of the profile in the Svetitsky--Yaffe mapping}
\label{app:moments_SY}

Let us consider the distribution $p(y) = \rho(y) / \int \rho(y) \, \diff y$, obtained normalizing the expression in Eq.~\eqref{eq:SY_profile}, for a given value of $R$. We managed to reduce the moments of this distribution in terms of special functions present in the \emph{scipy} python package, which allows practical computation of the width and the cumulants up to machine precision, given the mass scale $m$ and the distance $R$.

Let us start from equation \ref{eq:SY_profile}, neglecting some pre-factors, since they are cancelled by the normalization of the distribution. The computation of moments of the distribution is, thus, reduced to integrals of the following kind:
\begin{equation}
    I_{2n} = \int_{-\infty}^{+\infty} \frac{y^{2n}}{4y^2 + R^2} e^{-\sqrt{y^2 + R^2 / 4} / \lambda} \, \diff y=
    \frac{1}{4} \, r^{2n - 1} \int_{-\infty}^{+\infty} \diff \eta \, \sinh(\eta)^{2n} \, \frac{e^{-r \cosh(\eta) / \lambda}}{\cosh(\eta)}.
\end{equation}
We changed the integration variable to $\eta$, such that $\sinh(\eta) = 2 y / R$ and set $R / 2 = r$, for convenience. This integral can be written as a sum of $n + 1$ terms each involving an integral of the form
\begin{multline}
    \int_{-\infty}^{+\infty} \diff \eta \, \cosh(\eta)^{2k} \, \frac{e^{-r \cosh(\eta) / \lambda}}{\cosh(\eta)} = 
    \left. -\left(\frac{\partial}{\partial\alpha}\right)^{2k - 1} \int_{-\infty}^{+\infty} \diff \eta \, e^{-\alpha \cosh(\eta)} \right|_{\alpha = r / \lambda} = \\
    = \left. -2 \left(\frac{\partial}{\partial\alpha}\right)^{2n - 1} K_0(\alpha) \right|_{\alpha = r / \lambda},
\end{multline}
for $k = 0, \dots, n$. The special case $k = 0$ is to be intended as $-2$ times the primitive of $K_0$ that vanishes as its argument goes to infinity:
\begin{equation}
    -2 \left(\frac{\partial}{\partial \alpha}\right)^{-1} K_0(\alpha) \equiv
    2 \int_\alpha^\infty K_0(t) \diff t =
    \pi \left[1 - \alpha \, K_0(\alpha) \, L_{-1}(\alpha) - \alpha \, K_1(\alpha) \, L_0(\alpha)\right],
\end{equation}
where $L_i(\alpha)$ are the modified Struve functions.

Now we can go backwards and express the original integrals as
\begin{equation}
    I_{2n} = -\frac{1}{2} r^{2n-1} \sum_{k = 0}^n \binom{n}{k} (-1)^{n-k} \left. \left(\frac{\partial}{\partial\alpha}\right)^{2k - 1} K_0(\alpha) \right|_{\alpha = r / \lambda}
\end{equation}
and finally write the moments of the distribution as
\begin{equation}
    \mu_{2n} = I_{2n} / I_0
\end{equation}

From here, expanding the $K_0$ function and its (anti-)derivatives for large values of their arguments, we can obtain the large distance behaviour of the moments of the profile, according to the SY-mapping prediction. For example, the effective width of the flux tube results to be:
\begin{equation}
    w^2 = \frac{\lambda \, R}{2} - \frac{\lambda^2}{2} + \mathcal{O}\left( \frac{\lambda^3}{R} \right)
\end{equation}
and the Binder cumulant:
\begin{equation}
    b = -\frac{2 \, \lambda}{R} + \mathcal{O}\left( \frac{\lambda^2}{R^2} \right)
\end{equation}

\section{Overview of results from EST flow-based simulations}
\label{app:SNF}
 
In  a set of recent numerical studies the Nambu-Got\=o action and its higher order corrections were investigated through numerical simulations by directly regularizing the action on a bidimensional lattice~\cite{Caselle:2023mvh,Caselle:2024ent}. These simulations would be extremely challenging, if not unfeasible, with standard Markov Chain Monte Carlo methods; their realization has been made possible by the introduction of advanced deep-learning algorithms based on Normalizing Flows~\cite{rezende2015variational,Albergo:2019eim,Nicoli:2020njz,Cranmer:2023xbe}. In particular, these studies focused on the numerical determination of the the width and shape of the flux tube which one obtains assuming the \NambuG action (and its higher-order corrections up to the $\mathcal{K}^4$ term).  These predictions are too difficult to be obtained analytically and for this reason the authors of Refs.~\cite{Caselle:2023mvh,Caselle:2024ent} resorted to a numerical estimate. In the first proof-of-concept calculation~\cite{Caselle:2023mvh}, the authors found a perfect agreement with the predictions of Eqs.~\eqref{eq:EST_logbroad} and~\eqref{eq:EST_linbroad} in the limit of large $\sigma$. 
In the second work~\cite{Caselle:2024ent}, leveraging a powerful class of algorithms called Stochastic Normalizing Flows~\cite{wu2020stochastic,Caselle:2022acb,Bulgarelli:2024brv,Bonanno:2025pdp}, the full non-perturbative behaviour of Eq.~\eqref{eq:EST_linbroad} was confirmed numerically.

Furthermore, numerical results for the Binder cumulant associated with the string shape showed that the profile of the \NambuG contribution and its first higher-order correction is purely Gaussian: thus, the EST is width-less and possesses no intrinsic thickness. This result was further supported by the direct evaluation of the parameter $R_0$ (defined in Eq.~\eqref{eq:EST_logbroad}), which, in the large string tension regime, turned out to be always smaller than the lattice spacing of the regularized EST, thereby implying a widthless behaviour.

\FloatBarrier
\newpage

\section{Fit results at low and high temperatures}

\begin{table}[ht!]
    \centering
    \textbf{Exponential tails: fit details} \\ [0.5em]
    \begin{tabular}{|c|c|c|c|c|c|c|}
    \hline
    $\beta$ & $N_t$ & $R / a$ & $y_\intr$ & $A^{\text{(exp)}} \times 10^3$ & $\lambda / a$ & $\chi^2 / \ndof$\\
    \hline
    \multirow{4}{*}{8.768} & \multirow{4}{*}{24}
                    & 9  & 4 & 1.70567(14) & 1.612(63) & 2.17  \\
            &       & 11 & 5 & 1.60108(34) & 1.77(15)  & 1.13  \\
            &       & 13 & 5 & 3.05632(68) & 1.47(11)  & 0.58 \\
            &       & 15 & 5 & 2.20474(53) & 1.69(15)  & 2.96  \\  \hline
    \multirow{9}{*}{10.865}
            & 60    & 9  & 5 & 0.65(21)    & 1.91(26)  & 0.46 \\
            \cline{2-7}
            & \multirow{4}{*}{30}
                    & 9  & 6 & 0.74437(12) & 1.84(10) & 1.20 \\
            &       & 11 & 7 & 1.32587(75) & 1.67(25) & 1.12 \\
            &       & 13 & 7 & 0.78761(26) & 2.06(23) & 0.88 \\
            &       & 15 & 7 & 0.86927(40) & 2.08(32) & 0.39 \\
            \cline{2-7}
            & \multirow{4}{*}{20}
                    & 9  & 6 & 0.78876(14) & 1.79(11) & 1.81 \\
            &       & 11 & 7 & 0.92798(37) & 1.82(21) & 1.10 \\
            &       & 15 & 8 & 1.5923(11)  & 1.75(29) & 2.07 \\
            &       & 19 & 8 & 1.8404(14)  & 1.81(34) & 1.07 \\ \hline
    \multirow{4}{*}{12.963} & \multirow{4}{*}{36}
                    & 11 & 8 & 0.38802(28) & 2.23(48) & 1.64 \\ 
            &       & 13 & 8 & 0.28122(14) & 2.51(42) & 1.23 \\
            &       & 15 & 9 & 1.00678(43) & 1.86(20) & 1.54 \\
            &       & 17 & 9 & 0.33736(18) & 2.75(49) & 1.73 \\ \hline
    \end{tabular}
    \caption{Details of the fit results of the exponential tails of the profile at low temperatures as discussed in Section~\ref{sec:fit_exp}. The fit were performed taking into account the data from the transverse distance $y_\intr$ reported in the table, including it, up to the first value compatible with zero, excluding it, and the symmetric values, with respect to $y = 0$.}
    \label{tab:exp_fit_details}
\end{table}

\begin{table}[ht!]
	\centering
	\textbf{Convolution with $\sech$: fit details} \\ [0.5em]
	\begin{tabular}{|c|c|c|c|c|c|c|}
		\hline
		$\beta$ & $N_t$ & $R / a$ & $A^{\text{(conv)}}_1 \times 10^5$ & $\lambda / a$ & $s / a$ & $\chi^2 / \ndof$\\
		\hline
		\multirow{4}{*}{8.768} & \multirow{4}{*}{24}
		&  9 &  25.1(1.1) & 1.596(55) & 1.127(77) & 1.87  	\\ 
		&& 11 & 19.58(88) & 1.776(70) & 1.294(97) & 0.96	\\
		&& 13 & 15.32(23) & 1.631(87) & 1.764(93) & 1.04	\\
		&& 15 & 13.04(19) & 1.60(12) & 2.05(11) & 0.94  	\\  \hline
		\multirow{9}{*}{10.865}
		& 60    & 9 & 18.5(5.5) & 2.036(77) & 0.61(20) & 1.07 \\ \cline{2-7}
		& \multirow{4}{*}{30}
		&  9 &  22.2(2.5) & 2.041(22) & 0.515(61) & 2.87   	\\
		&& 11 &  9.13(36) & 2.111(39) & 1.193(62) & 1.36 	\\
		&& 13 &  6.71(31) & 2.228(83) & 1.53(11) & 1.01  	\\
		&& 15 &  5.24(26) & 2.31(14) & 1.90(17) & 0.51   	\\ \cline{2-7}
		& \multirow{4}{*}{20}
		&  9 &  14.44(59) & 1.931(22) & 0.833(40) & 3.43 	\\
		&& 11 &  8.32(32) & 2.037(57) & 1.322(78) & 1.20	\\
		&& 15 &  4.944(79) & 2.01(14) & 2.17(16) & 1.57 	\\
		&& 19 &  3.72(11) & 2.07(36) & 2.68(36) & 1.64  	\\ \hline
		\multirow{4}{*}{12.963} & \multirow{4}{*}{36}
		 & 11 &     5.73(80) & 2.368(80) & 0.98(16) & 1.69	\\ 
		&& 13 &  3.19(17) & 2.350(99) & 1.71(14) & 1.02	\\
		&& 15 &  2.73(17) & 2.50(14) & 1.88(19) & 1.90 	\\
		&& 17 &  2.75(39) & 3.04(17) & 1.57(28) & 1.48 	\\
		 \hline
	\end{tabular}
	\caption{Details of the results for fits performed at low temperatures assuming the $\sech$ convolution model of Section~\ref{sec:fit_convolutions} using Eq.~\eqref{eq:conv_sech}. All the data not compatible with zero are included. For comparison, the values of $1 / M_0$ (interpolated from Ref.~\cite{Teper:1998te}) are $1.294(12)$ at $\beta = 8.768$, $1.595(17)$ at $\beta = 10.865$ and $1.938(23)$ at $\beta = 12.9625$.}
	\label{tab:conv_fit_details}
\end{table}

\begin{table}[ht!]
    \centering
    \textbf{Convolution with $\exp$: fit details} \\ [0.5em]
    \begin{tabular}{|c|c|c|c|c|c|c|}
    \hline
    $\beta$ & $N_t$ & $R / a$ & $A^{\text{(conv)}}_2 \times 10^4$ & $\lambda / a$ & $s / a$ & $\chi^2 / \ndof$\\
    \hline
    \multirow{4}{*}{8.768} & \multirow{4}{*}{24}
                    & 9 &  8.51(69) & 1.626(60) & 1.526(37) & 1.96 \\
            &       & 11 & 7.52(72) & 1.831(79) & 1.713(50) & 0.97 \\ 
            &       & 13 & 35(29) & 1.27(20) & 2.31(11) & 1.08 \\
            &       & 15 & 300(1400) & 1.01(46) & 2.72(23) & 1.65 \\ \hline
    \multirow{9}{*}{10.865}
            & 60    & 9 & 2.63(17) & 2.146(96) & 1.390(58) & 1.16 \\ \cline{2-7}
            & \multirow{4}{*}{30}
                    & 9 &  2.633(45) & 2.143(28) & 1.355(15) & 4.80 \\
            &       & 11 & 2.90(11) & 2.187(45) & 1.783(28) & 1.64 \\
            &       & 13 & 2.95(23) & 2.298(93) & 2.077(50) & 1.04 \\
            &       & 15 & 3.18(48) & 2.38(15) & 2.399(96) & 0.52  \\ \cline{2-7}
            & \multirow{4}{*}{20}
                    & 9 &  2.999(61) & 2.006(26) & 1.467(14) & 4.36 \\
            &       & 11 & 3.14(19) & 2.087(64) & 1.852(35) & 1.41 \\
            &       & 15 & 4.5(1.1) & 2.04(15) & 2.535(98) & 1.60   \\
            &       & 19 & 5.5(3.8) & 2.09(38) & 3.01(23) & 1.64    \\ \hline
    \multirow{4}{*}{12.963} & \multirow{4}{*}{36}
                    & 11 & 1.378(85) & 2.459(91) & 1.768(56) & 1.70 \\ 
            &       & 13 & 1.63(16) & 2.41(11) & 2.269(74) & 1.03   \\
            &       & 15 & 1.58(21) & 2.55(15) & 2.469(97) & 1.93   \\
            &       & 17 & 1.13(12) & 3.14(19) & 2.48(12) & 1.49    \\ \hline
    \end{tabular}
    \caption{Details of the results for fits performed at low temperatures assuming the convolution model of Section~\ref{sec:fit_convolutions} using Eq.~\eqref{eq:conv}. All the data not compatible with zero are included.}
    \label{tab:conv_fit_details2}
\end{table}

\begin{table}[ht!]
    \centering
    \textbf{Clem: fit details} \\ [0.5em]
    \begin{tabular}{|c|c|c|c|c|c|c|}
    \hline
    $\beta$ & $N_t$ & $R / a$ & $A^{\text{(Clem)}}$ & $\lambda / a$ & $\xi / a$ & $\chi^2 / \ndof$\\
    \hline
    \multirow{4}{*}{8.768} & \multirow{4}{*}{24}
                    & 9 & 0.0110(37) & 1.406(81) & 3.69(20) & 1.69 \\
            &       & 11 & 0.0145(66) & 1.49(10) & 4.42(29) & 0.99 \\
            &       & 13 & 0.014(13)  & 1.43(10) & 6.20(51) & 1.11 \\
            &       & 15 & 0.06(9)    & 1.36(13) & 7.56(82) & 0.90 \\  \hline
    \multirow{9}{*}{10.865}
            & 60    & 9 & 0.00145(42) & 1.95(15) & 3.13(22) & 1.02 \\ \cline{2-7}
            & \multirow{4}{*}{30}
                    & 9 & 0.00159(11) & 1.886(34) & 3.134(48) & 2.12 \\
            &       & 11 & 0.00314(55) & 1.857(62) & 4.32(14) & 0.91 \\
            &       & 13 & 0.0091(37) & 1.707(95) & 5.70(30) & 1.22  \\
            &       & 15 & 0.0105(76) & 1.81(17) & 6.42(59) & 0.85   \\ \cline{2-7}
            & \multirow{4}{*}{20}
                    & 9 & 0.00203(18) & 1.801(37) & 3.335(60) & 2.59 \\
            &       & 11 & 0.0041(11) & 1.768(83) & 4.55(19) & 0.79 \\
            &       & 15 & 0.010(10) & 1.83(26) & 6.39(81) & 0.86   \\
            &       & 19 & 0.014(20) & 1.96(34) & 7.7(1.3) & 1.45   \\ \hline
    \multirow{4}{*}{12.963} & \multirow{4}{*}{36}
                    & 11 & 0.00075(20) & 2.32(16) & 3.86(24) & 0.86 \\ 
            &       & 13 & 0.0029(18) & 1.98(19) & 5.77(52) & 1.33  \\
            &       & 15 & 0.0020(13) & 2.28(24) & 6.04(60) & 1.16  \\
            &       & 17 & 0.0020(16) & 2.44(31) & 6.61(79) & 1.03  \\ \hline
    \end{tabular}
    \caption{Details of the results for fits performed at low temperatures assuming the Clem model discussed in Sections~\ref{sec:clem} and~\ref{sec:fit_clem} using Eq.~\eqref{eq:clem}. All the data not compatible with zero were included.}
    \label{tab:clem_fit_details}
\end{table}


\begin{table}[ht!]
    \centering
    \textbf{High $T$: fit details for $\beta=10.865$ and $11.914$} \\ [0.5em]
    \begin{tabular}{|c|c|c|c|c|c|}
    \hline
    $\beta$ & $N_t$ & $R / a$ & $A^{\text{(SY)}}$ & $\lambda / a$ & $\chi^2 / \ndof$\\ \hline
    \multirow{18}{*}{10.865} & \multirow{8}{*}{10}
    &    11 & 0.0002785(47) & 4.62(16) & 0.96 \\ 
    &&   15 & 0.0002726(82) & 4.59(28) & 0.70 \\ 
    &&   17 & 0.000271(11)  & 4.66(40) & 1.25 \\ 
    &&   19 & 0.000286(18)  & 5.41(73) & 0.51 \\ 
    &&   21 & 0.000276(24)  & 5.25(96) & 0.65 \\ \cline{3-6}
    && $> 10$ & 0.0002777(43) & 4.63(14) & 0.81 \\
    && $> 12$ & 0.0002768(59) & 4.70(20) & 0.82 \\ \cline{3-6}
    && \multicolumn{2}{|l|}{From $\braket{P \, P^\dagger}$}
                              & 4.5192(98) & \\ \cline{2-6}
                            & \multirow{10}{*}{8}
     & 9 & 0.0003387(37) & 9.35(24) & 3.77 \\ 
    && 11 & 0.0003371(45) & 9.90(31) & 1.93 \\ 
    && 13 & 0.0003366(53) & 10.17(38) & 0.98 \\ 
    && 15 & 0.0003314(63) & 10.53(47) & 1.01 \\ 
    && 17 & 0.0003195(69) & 9.94(50) & 1.11 \\ 
    && 19 & 0.0003242(84) & 10.92(67) & 1.24 \\ 
    && 21 & 0.000332(11) & 12.33(95) & 1.18 \\  \cline{3-6}
    && $ > 11 $ & 0.0003442(42) & 10.61(30) & 1.63 \\
    && $ > 13 $ & 0.0003419(50) & 10.87(37) & 1.46 \\ \cline{3-6}
    && \multicolumn{2}{|l|}{From $\braket{P \, P^\dagger}$}
                                & 10.083(30) & \\ \hline
    \multirow{8}{*}{11.914} & \multirow{8}{*}{11}
     & 11 & 0.0002188(40) & 5.24(20) & 0.65 \\ 
    && 13 & 0.0002214(56) & 5.57(30) & 0.83 \\ 
    && 15 & 0.0002246(76) & 5.87(44) & 0.82 \\ 
    && 17 & 0.000229(10) & 6.14(62) & 0.77 \\ 
    && 19 & 0.000226(14) & 6.23(86) & 0.98 \\ 
    && 21 & 0.000215(16) & 5.40(89) & 1.36 \\ \cline{3-6}
    && $ > 11 $ & 0.0002215(37) & 5.44(19) & 0.87 \\
    && $ > 13 $ & 0.0002231(50) & 5.71(27) & 0.85 \\ \cline{3-6}
    && \multicolumn{2}{|l|}{From $\braket{P \, P^\dagger}$}
                                & 5.053(11) & \\ \hline
    \end{tabular}
    \caption{Details of the results for fits performed at high temperatures assuming the model based on the Svetitsky--Yaffe mapping for $\beta=10.865$ and $11.914$. We also reported the results for combined fits using all the data with $R / a$ greater than 11 and 13 lattice spacings and the value of $\lambda$ that we extracted from the Polyakov loops correlator, which is $1 / (2 \, E_0)$ if $\braket{P(0) \, P^\dagger(R)} \sim \exp(-E_0 \, R)$.}
    \label{tab:symp_fit_details}
\end{table}

\begin{table}[ht!]
    \centering
    \textbf{High $T$: fit details for $\beta=12.962$ and $14.011$} \\ [0.5em]
    \begin{tabular}{|c|c|c|c|c|c|}
    \hline
    $\beta$ & $N_t$ & $R / a$ & $A^{\text{ (SY)}}$ & $\lambda / a$ & $\chi^2 / \ndof$\\ \hline
    \multirow{17}{*}{12.962} & \multirow{6}{*}{12}
     & 11 & 0.0001701(24) & 5.40(16) & 1.62 \\ 
    && 13 & 0.0001669(31) & 5.46(21) & 0.81 \\ 
    && 15 & 0.0001676(41) & 5.80(30) & 1.12 \\ 
    && 17 & 0.0001588(48) & 5.26(33) & 0.59 \\ \cline{3-6}
    && $ > 11 $ & 0.0001701(22) & 5.51(15) & 1.35 \\
    && $ > 13 $ & 0.0001664(29) & 5.51(20) & 0.97 \\ \cline{3-6}
    && \multicolumn{2}{|l|}{From $\braket{P \, P^\dagger}$}
                                & 5.534(26) & \\ \cline{2-6}
    &                         \multirow{11}{*}{10}
     &  9 & 0.0001745(25) & 9.07(31) & 2.83 \\ 
    && 11 & 0.0001803(34) & 11.05(50) & 1.89 \\ 
    && 13 & 0.0001853(44) & 13.05(76) & 1.46 \\ 
    && 15 & 0.0001818(50) & 13.25(90) & 1.15 \\ 
    && 17 & 0.0001846(60) & 14.5(1.2) & 1.12 \\ 
    && 19 & 0.0001819(67) & 14.5(1.3) & 0.93 \\ 
    && 21 & 0.0001735(70) & 13.2(1.3) & 1.31 \\ \cline{3-6}
    && $ > 11 $ & 0.0001900(34) & 13.07(56) & 1.75 \\
    && $ > 13 $ & 0.0001882(41) & 13.97(74) & 1.40 \\ \cline{3-6}
    && \multicolumn{2}{|l|}{From $\braket{P \, P^\dagger}$}
                                & 12.30(16) & \\ \hline
    \multirow{10}{*}{14.011} & \multirow{10}{*}{13}
     &  9 & 0.0001323(23) & 5.06(19) & 1.32 \\ 
    && 11 & 0.0001336(32) & 5.73(30) & 0.62 \\ 
    && 13 & 0.0001282(39) & 5.40(35) & 1.08 \\ 
    && 15 & 0.0001216(47) & 5.00(40) & 0.67 \\ 
    && 17 & 0.0001150(57) & 4.76(49) & 0.91 \\ 
    && 19 & 0.0001195(76) & 5.43(73) & 1.26 \\ 
    && 21 & 0.0001074(93) & 4.77(86) & 0.82 \\ \cline{3-6}
    && $ > 11 $ & 0.0001308(28) & 5.57(25) & 1.01 \\ 
    && $ > 13 $ & 0.0001253(34) & 5.27(30) & 1.06 \\ \cline{3-6}
    && \multicolumn{2}{|l|}{From $\braket{P \, P^\dagger}$} & 6.039(36) & \\ \hline
    \end{tabular}
    \caption{Same as Tab.~\ref{tab:symp_fit_details}, but for fits performed on data collected at $\beta=12.962$ and $14.011$.}
    \label{tab:symp_fit_details2}
\end{table}

\FloatBarrier

\providecommand{\href}[2]{#2}\begingroup\raggedright\endgroup

\end{document}

%% file: poly_poly_plaq.tex
\begin{tikzpicture}[scale=1.1, transform shape]

    \draw [ -> ] (-1, 0) -- (-1, 1) node [pos=0.95, anchor=south] {$\hat 0$};
    \draw [ -> ] (-1, 0) -- (0, 0) node [pos=0.95, anchor=north] {$\hat 1$};
    \draw [ -> ] (-1, 0) -- (-0.5, 0.5) node [pos=0.95, anchor=south] {$\hat 2$};

    \draw [red, ultra thick, ->] (0.5, 0) -- (0.5, 1.5) node [anchor=south east] {$P(0, 0)$};
    \draw [red, ultra thick] (0.5, 1.3) -- (0.5, 3.5);

    \draw [blue, ultra thick] (7.5, 0) -- (7.5, 1.5) node [anchor=north west] {$P^\dagger(R, 0)$};
    \draw [blue, ultra thick, ->] (7.5, 3.5) -- (7.5, 1.3);

    \draw [ultra thin] (0.5, 1.5) -- (7.5, 1.5);
    \draw [ultra thin] (1.0, 2.0) -- (8.0, 2.0);
    \draw [ultra thin] (1.5, 2.5) -- (8.5, 2.5);

    \draw [ultra thin] (0.5, 1.5) -- (1.5, 2.5);
    \draw [ultra thin] (1.5, 1.5) -- (2.5, 2.5);
    \draw [ultra thin] (2.5, 1.5) -- (3.5, 2.5);
    \draw [ultra thin] (3.5, 1.5) -- (4.5, 2.5);
    \draw [ultra thin] (4.5, 1.5) -- (5.5, 2.5);
    \draw [ultra thin] (5.5, 1.5) -- (6.5, 2.5);
    \draw [ultra thin] (6.5, 1.5) -- (7.5, 2.5);
    \draw [ultra thin] (7.5, 1.5) -- (8.5, 2.5);

    \draw [ultra thick] (4.5, 2.5) -- (5.5, 2.5) -- (5.5, 3.5) -- (4.5, 3.5) -- cycle;
    \node at (4.5, 3) [anchor=east] {$\Pi_{01} \left( \frac{R - a}{2}, y \right)$};

    \draw [ <-> ] (0.6, 0.5) -- (7.4, 0.5) node [pos=0.5, anchor=north] {$R$};
    \draw [ <-> ] (0.6, 1.35) -- (3.4, 1.35) node [pos=0.5, anchor=north] {$(R - a) / 2$};
    \draw [ <-> ] (3.65, 1.55) -- (4.5, 2.4) node [pos=0.5, anchor=north] {$y$};

\end{tikzpicture}